\documentclass[oldversion]{aa}
\usepackage{graphicx}
\usepackage{natbib}
\usepackage{amsmath}
\usepackage{txfonts}
\usepackage{url}

\newcommand{\lsim}{\lesssim}

\bibpunct{(}{)}{;}{a}{}{,}

\begin{document}
\title{ESO Imaging Survey: Optical follow-up of 12 selected XMM-Newton
  fields\thanks{Based on observations carried out at the European
    Southern Observatory, La Silla, Chile under program Nos.
    170.A-0789, 70.A-0529, 71.A-0110.}\fnmsep\thanks{Appendices
    \ref{sec:field_desc} to \ref{sec:catalogs} and Fig.~\ref{fig:xray}
    are only available in electronic form at
    \texttt{http://www.edpsciences.org}}}
\author{J.\,P.~Dietrich\inst{1,2} \and J.-M.~Miralles\inst{2} \and
  L.\,F.~Olsen\inst{2,3,4} \and L.~da~Costa\inst{2} \and
  A.~Schwope\inst{5} \and C.~Benoist\inst{2,4} \and
  V.~Hambaryan\inst{5} \and A.~Mignano\inst{6,2,7} \and
  C.~Motch\inst{8} \and C.~Rit\'e\inst{2} \and R.~Slijkhuis\inst{2}
  \and J.~Tedds\inst{9} \and B.~Vandame\inst{2} \and
  M.\,G.~Watson\inst{9} \and S.~Zaggia\inst{2,10}}

\offprints{J.~P.~Dietrich}
\institute{Institut f\"ur Astrophysik und extraterrestrische
  Forschung, University of Bonn, Auf dem H\"ugel 71, 53121 Bonn, 
  Germany \\\email{dietrich@astro.uni-bonn.de}
  \and
  European Southern Observatory, Karl-Schwarzschild-Str.~2, 85748
  Garching b. M\"unchen, Germany
  \and
  Copenhagen University Observatory, Juliane Maries Vej~30, DK-2100 
  Copenhagen, Denmark
  \and
  Observatoire de la C\^ote d'Azur, Laboratoire
  Cassiop\'ee, BP4229, 06304 Nice Cedex~4, France
  \and
  Astrophysikalisches Institut Potsdam, An der Sternwarte~16, 14482
  Potsdam, Germany
  \and
  Dipartimento di Astronomia, Universit\`a di Bologna, via Ranzani~1, I-40126
  Bologna, Italy
  \and
  Istituto di Radioastronomia, INAF, Via Gobetti 101, I-40129 Bologna,
  Italy
  \and
  Observatoire Astronomique, CNRS UMR 7550, 11~rue de l'Universit\'e,
  67000 Strasbourg, France
  \and
  Department of Physics and Astronomy, University of Leicester,
  Leicester LE1~7RH, UK
  \and
  Osservatorio Astronomico di Trieste, Via G.\,B. Tiepolo, 11, I-34131
  Trieste, Italy
}

\date{Received 7 July 2005; Accepted 7 October 2005}

\abstract{This paper presents the data recently released for the
  XMM-Newton/WFI survey carried out as part of the ESO Imaging Survey
  (EIS) project. The aim of this survey is to provide optical imaging
  follow-up data in $BVRI$ for identification of serendipitously
  detected X-ray sources in selected XMM-Newton fields. In this paper, fully
  calibrated individual and stacked images of 12 fields as well as
  science-grade catalogs for the 8 fields located at high-galactic
  latitude are presented. These products were created, calibrated and
  released using the infrastructure provided by the EIS Data Reduction
  system and its associated EIS/MVM image processing engine, both of
  which are briefly described here. The data covers an area of $\sim
  3$~square degrees for each of the four passbands. The median seeing
  as measured in the final stacked images is $0\farcs94$, ranging from
  $0\farcs60$ and $1\farcs51$. The median limiting magnitudes (AB
  system, 2\arcsec\ aperture, $5\sigma$ detection limit) are 25.20,
  24.92, 24.66, and 24.39~mag for $B$-, $V$-, $R$-, and $I$-band,
  respectively. When only the 8 high-galactic latitude fields are
  included these become 25.33, 25.05, 25.36, and 24.58~mag, in good
  agreement with the planned depth of the survey. Visual inspection of
  images and catalogs, comparison of statistics derived from the
  present data with those obtained by other authors and model
  predictions, as well as direct comparison of the results obtained
  from independent reductions of the same data, demonstrate the
  science-grade quality of the automatically produced final images and
  catalogs. These survey products, together with their logs, are
  available to the community for science exploitation in conjunction
  with their X-ray counterparts. Preliminary results from the
  X-ray/optical cross-correlation analysis show that about 61\% of the
  detected X-ray point sources in deep XMM-Newton exposures have at least one
  optical counterpart within 2\arcsec\ radius down to $R \simeq$
  25~mag, 50\% of which are so faint as to require VLT observations
  thereby meeting one of the top requirements of the survey, namely to
  produce large samples for spectroscopic follow-up with the VLT,
  whereas only 15\% of the objects have counterparts down to the DSS
  limiting magnitude.

} 
\maketitle

\section{Introduction}
\label{sec:introduction}
The new generation of highly sensitive X-ray observatories such as
Chandra and XMM-Newton is generating large volumes of X-ray data,
which through public archives are made available for all
researchers. Even though all observations are targeting a particular
object, the large field of view (FOV) of XMM-Newton allows many other
sources to be detected in deep exposures. These sources are the main
product of the XMM-Newton Serendipitous Sky Survey
\citep{2001A&A...365L..51W}, which annually identifies about 50\,000
new X-ray sources. To fully understand the nature of these
serendipitously detected sources follow-up observations at other
wavelengths are needed.

\begin{table*}
  \centering
  \caption{Central positions in right ascension, declination and
    galactic longitude and latitude of the 12 XMM-Newton fields
    observed as part of the XMM-Newton follow-up survey.}
  \begin{tabular}{llrrrr}
    \hline\hline
    Field & Target & $\alpha$ (J2000.0) & $\delta$ (J2000.0) &
    \multicolumn{1}{c}{$l$} & \multicolumn{1}{c}{$b$} \\\hline
    XMM-01 & RX J0925.7$-$4758 & 09:25:46.0 & $-$47:58:17 & 271:21:18 &
    +01:53:03\\
    XMM-02 & RX J0720.4$-$3125 & 07:20:25.1 & $-$31:25:49 & 244:09:28 &
    $-$08:09:50 \\
    XMM-03 & HE 1104$-$1805 & 11:06:33.0 & $-$18:21:24 & 270:49:55 &
    +37:53:29 \\
    XMM-04 & MS 1054.4$-$0321 & 10:56:60.0 & $-$03:37:27 & 256:34:30 &
    +48:40:18 \\
    XMM-05 & BPM 16274 & 00:50:03.2 & $-$52:08:17 & 303:26:03 &
    $-$64:59:19 \\
    XMM-06 & RX J0505.3$-$2849 & 05:05:20.0 & $-$28:49:05 & 230:39:29 &
    $-$34:36:50 \\
    XMM-07 & LBQS 2212$-$1759 & 22:15:31.7 & $-$17:44:05 &  39:16:07 &
    $-$ 52:55:44\\
    XMM-08 & NGC 4666 & 12:45:08.9 & $-$00:27:38 & 299:25:55 &
    +63:17:22 \\
    XMM-09 & QSO B1246$-$057 & 12:49:13.9 & $-$05:59:19 & 301:55:40 &
    +56:52:43 \\
    XMM-10 & PB 5062 & 22:05:09.8 & $-$01:55:18 & 58:03:55 &
    $-$42:54:13 \\
    XMM-11 & Sgr A & 17:45:40.0 & $-$29:00:28 & 359:56:39 &
    $-$00:02:45 \\
    XMM-12 & WR 46 & 12:05:19.0& $-$62:03:07 & 297:33:23 & +00:20:14
    \\ 
    \hline
  \end{tabular}
  \label{tab:field-coord}
\end{table*}

Based on a Call for Ideas for public surveys to the ESO community, the
XMM-Newton Survey Science Center (SSC) proposed optical follow-up
observations of XMM-Newton fields for its X-ray Identification (XID)
program
\citep{2001A&A...365L..51W,2002A&A...382..522B,2004A&A...428..383D}.
This proposal was evaluated and accepted by ESO's Survey Working Group
(SWG) and turned into a proposal for an ESO large program submitted to
the ESO OPC.\footnote{The full text of the large program proposal is
available at
\url{http://www.eso.org/science/eis/documents/EIS.2002-09-04T12:42:31.890.ps.gz}}

The XMM-Newton optical follow-up survey aims at obtaining optical
observations of XMM-Newton Serendipitous Sky Survey fields, publicly
available in the XMM-Newton archive, using the wide-field imager (WFI) at the
ESO/MPG 2.2m telescope at the La Silla Observatory. WFI has a FOV
which is an excellent match to that of the X-ray detectors on-board
the XMM-Newton satellite, making this instrument an obvious choice for
this survey in the South. A complementary multiband optical imaging
program (to median $5\sigma$ limiting magnitudes reaching
$i^\prime=23.1$) for over 150 XMM-Newton fields is nearing completion in the
North using the similarly well matched Wide Field Camera on the 2.5~m
Isaac Newton Telescope
\citep{2003AN....324..178Y,2003AN....324...89W}.  In order to provide
data for minimum spectral discrimination and photometric redshift
estimates of the optical counterparts of previously detected X-ray
sources, the survey has been carried out in the $B$-, $V$-, $R$-, and
$I$-passbands. The survey has been administered and carried out by the
ESO Imaging Survey (EIS) team.

This paper describes observations, reduction, and science verification
of data publicly released as part of this follow-up survey.
Section~\ref{sec:targets} briefly describes the X-ray observations while
Sect.~\ref{sec:observations} focus on the optical imaging. In
Sect.~\ref{sec:reduction} the reduction and calibration of optical
data are presented and the results discussed. Final survey products
such as stacked images and science-grade catalogs extracted from them
are presented in Sect.~\ref{sec:products}. The quality of these
products is evaluated in Sect.~\ref{sec:discussion} by comparing
statistical measures obtained from these data to those of other
authors as well as from a direct comparison with the results of an
independent reduction of the same dataset. In this section the results
of a preliminary assessment of X-ray/optical cross-correlation are
also discussed. Finally, in Sect.~\ref{sec:summary} a brief summary of
the paper is presented.

\section{X-ray observations}
\label{sec:targets}
The original proposal by the SWG to the ESO OPC was to cover a total
area of approximately 10 square degrees (40 fields) to a limiting
magnitude of 25 (AB, $5\sigma$, 2\arcsec\ aperture). The OPC approved
enough time to observe 12 fields, later extending the time allocation
to include 3 more fields. This paper presents results for the original
12 fields for which the optical data were originally publicly released
in the fall of 2004, with corrections to the weight maps
  released in July 2005. Table~\ref{tab:field-coord} gives the
location of the 12 fields listing: in Col.~1 the field name; in
Col.~2 the original XMM-Newton target name; in Cols.~3 and 4 the right
ascension and declination in J2000; and in Cols.~5 and 6 the galactic
coordinates, $l$ and $b$.

The 12 fields listed in Table~\ref{tab:field-coord} were selected and
prioritized by a collaboration of interested parties from the SSC, a
group at the Institut f\"ur Astrophysik und Extraterrestrische
Forschung (IAEF) of the University of Bonn, and an appointed committee
of the SWG. These fields were selected following, as much as possible,
the criteria given in the proposal, namely that: (1) the fields had to
have a large effective exposure time in X-ray (ideally $t_\mathrm{exp} >
30$~ks) with no enhanced background; (2) the X-ray data of the
selected fields had to be public by the time the raw WFI frames were
to become public; (3) the original targets should not be too bright
and/or extended, thus allowing a number of other X-ray sources to be
detected away from the primary target; and (4) $\sim 70$\% of the
fields had to be located at high-galactic latitude. Comments on the
individual fields can be found in Appendix~\ref{sec:field_desc}.

Combined EPIC X-ray images for the fields listed in
Table~\ref{tab:field-coord} were created from exposures taken with the
three cameras (PN, MOS1, MOS2) on-board XMM-Newton. The
  sensitive area of these cameras is a circle with a diameter of
  approximately 30\arcmin. The contributing
X-ray observations are summarized in Table~\ref{tab:xray_obs} which
gives for each field: in Col.~1 the field identification; in Col.~2
the XMM-Newton observation id; in Col.~3 the nominal exposure time; in
Cols.~4--6 the settings for each of the cameras. Here (E)FF indicates
(extended) full frame readout, LW large window mode and SW2 small
window mode. These cameras and their settings are described in detail
in \citet{2004.xmm.guide.E}. For some fields additional observations
were available but these were discarded mainly due to unsuitable
camera settings.

\begin{table*}
  \centering
  \caption{Information about X-ray imaging used to create composite
X-ray images.} 
  \begin{tabular}{llrrrr}
    \hline\hline
    Field & Obs. ID & $T_\mathrm{exp}$ (s)& \multicolumn{3}{c}{Camera
      settings}\\\hline
    XMM-01 & 0111150201 & 62\,067 & EPN LW & MOS1 FF & MOS2 SW2\\
           & 0111150101 & 61\,467 & EPN LW & MOS1 FF & MOS2 SW2\\
    \hline
    XMM-02 & 0164560501 & 50\,059 & EPN FF & MOS1 FF & MOS2 FF\\
           & 0156960201 & 30\,243 & EPN FF & MOS1 FF & MOS2 FF\\
           & 0156960401 & 32\,039 & EPN FF & MOS1 FF & MOS2 FF\\
    \hline
    XMM-03 & 0112630101 & 36\,428 & EPN FF & MOS1 FF & MOS2 FF\\
    \hline
    XMM-04 & 0094800101 & 41\,021 & EPN FF & MOS1 FF & MOS2 FF\\
    \hline
    XMM-05 & 0125320701 & 45\,951 & EPN FF & MOS1 FF & MOS2 FF\\ 
           & 0125320401 & 33\,728 & EPN FF & MOS1 FF & MOS2 FF\\
           & 0125320501 &  7845   & EPN FF & MOS1 FF & MOS2 FF\\
           & 0153950101 &  5156   & EPN FF & MOS1 FF & MOS2 FF\\
           & 0133120301 & 12\,022 & EPN FF & MOS1 FF & MOS2 FF\\
           & 0133120401 & 13\,707 & EPN FF & MOS1 FF & MOS2 FF\\
    \hline
    XMM-06 & 0111160201 & 49\,616 & EPN EFF &MOS1 FF & MOS2 FF\\ 
    \hline
    XMM-07 & 0106660501 &  11\,568 & EPN FF & MOS1 FF & MOS2 FF\\ 
           & 0106660401 &  35\,114 & ---       & MOS1 FF & MOS2 FF\\  
           & 0106660101 &  60\,508 & EPN FF & MOS1 FF & MOS2 FF\\ 
           & 0106660201 &  53\,769 & EPN FF & MOS1 FF & MOS2 FF\\ 
           & 0106660601 & 110\,168 & EPN FF & MOS1 FF & MOS2 FF\\ 
    \hline
    XMM-08 & 0110980201 &  58\,237 & EPN EFF & MOS1 FF & MOS2 FF\\
    \hline
    XMM-09 & 0060370201 &  41\,273 & EPN FF & MOS1 FF & MOS2 FF\\
    \hline
    XMM-10 & 0012440301 &  35\,366 & EPN FF & MOS1 FF & MOS2 FF\\
    \hline
    XMM-11 & 0112970601 & 27\,871 & EPN FF & ---        & ---       \\
           & 0112971601 & 28\,292 & ---       & MOS1 FF & MOS2 FF\\
           & 0112972101 & 26\,870 & EPN FF & MOS1 FF & MOS2 FF\\
           & 0111350301 & 17\,252 & EPN FF & MOS1 FF & MOS2 FF\\
           & 0111350101 & 52\,823 & EPN FF & MOS1 FF & MOS2 FF\\
    \hline
    XMM-12 & 0109110101 & 76\,625 & EPN EFF & MOS1 FF & MOS2 FF\\
    \hline
  \end{tabular}
  \label{tab:xray_obs}
\end{table*}

\begin{figure*}
\centering
\resizebox{\hsize}{!}{\includegraphics[width=17cm]{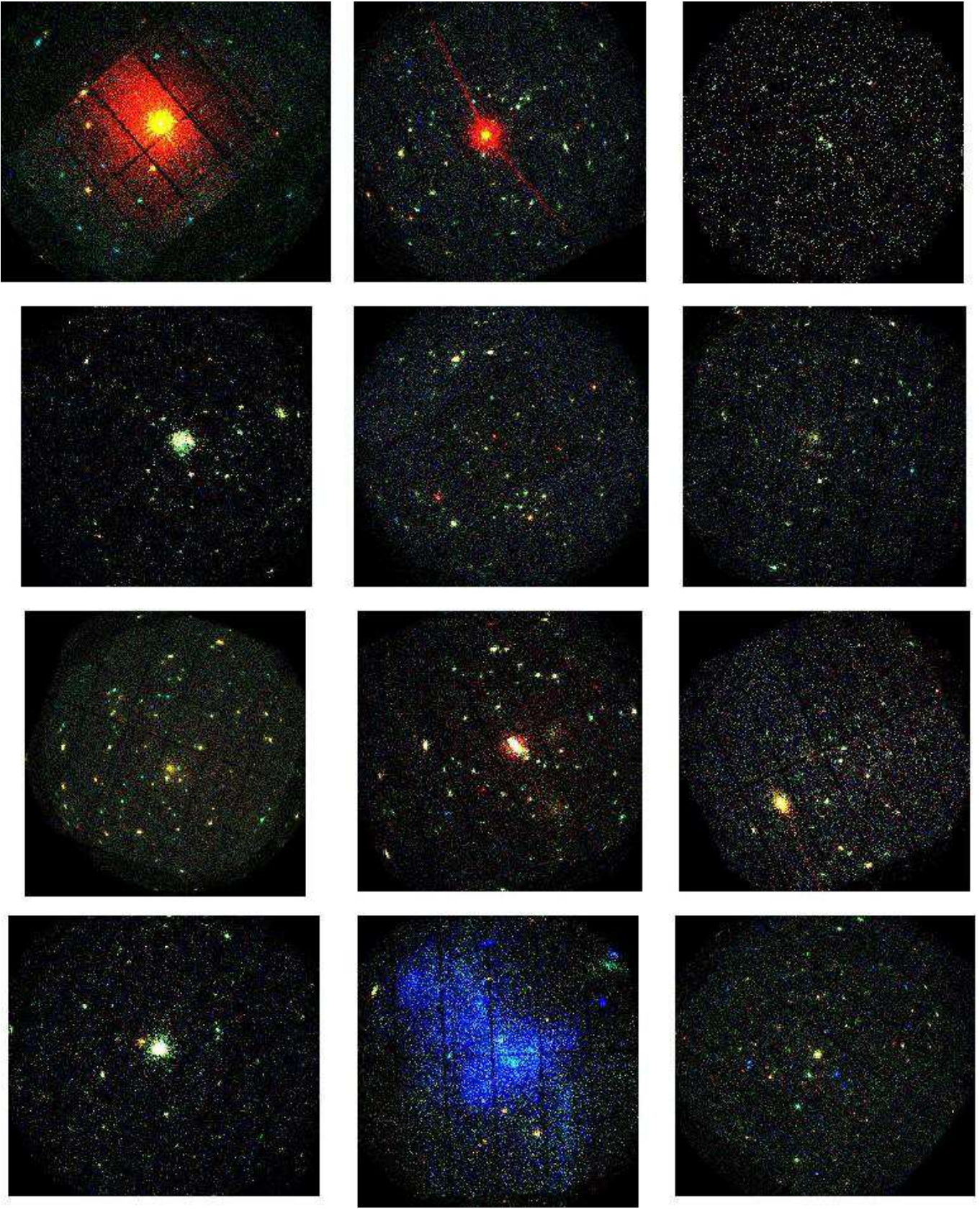}}
\caption{Color composite X-ray images for the 12 fields considered in
  this paper (XMM-01 to XMM-12 from top left to bottom right). The
  color images are composites within the so-called XID-band
  (0.5--4.5~keV). Red, green and blue channels comprise the energy
  ranges 0.5--1.0~keV, 1.0--2.0~keV, and 2.0--4.5~keV, respectively.
  Weighting of the sub-images was done in a manner that a typical
  extragalactic source with a power law spectrum with photon index 1.5
  and absorption column density $N_\mathrm{H} =1 \times
  10^{20}$\,cm$^{-2}$ would have equal photon numbers in all three
  bands. North is up and East to the left. The size of the images
  is typically $30\arcmin\times30\arcmin$ but varies slightly with
  camera orientation.} 
\label{fig:xray}
\end{figure*}

The XMM-Newton data, both in raw and pipeline reduced form, are
available through the XMM-Newton Science
Archive.\footnote{\url{http://xmm.vilspa.esa.es/external/xmm_data_acc/xsa/index.shtml}}
These data were used to create a wide range of products which include:
\begin{itemize}
\item Combined EPIC images in the XID-band 0.5--4.5 keV (FITS); 
\item Combined EPIC images in the total band 0.1--12 keV (FITS); 
\item Color images using three sub-bands, 0.5--1.0 keV (red), 1.0--2.0 keV
(green), 2.0--4.5 keV (blue), in the XID-band (JPG).
\end{itemize}

As an illustration, Fig.~\ref{fig:xray} shows color composites of
the final combined X-ray images for the 12 fields considered. Note
that the X-ray images have a non-uniform exposure time over the field
of view due to (1) the arrangements of the CCDs in the focal plane,
which is different for the three cameras, and (2) the vignetting of
the camera optics.

\section{Optical observations}
\label{sec:observations}
As mentioned earlier, the optical observations were carried out using
WFI at the ESO/MPG-2.2m telescope in service mode. WFI is a focal
reducer-type mosaic camera mounted at the Cassegrain focus of the
telescope. The mosaic consists of $4 \times 2$ CCD chips with $2048
\times 4096$ pixels with a projected pixel size of $0\farcs{238}$,
giving a FOV of $8\farcm{12} \times 16\farcm{25}$ for each individual
chip. The chips are separated by gaps of $23\farcs{8}$ and
$14\farcs{3}$ along the right ascension and declination direction,
respectively. The full FOV of WFI is thus $34\arcmin \times 33\arcmin$
with a filling factor of $95.9$\%.

The WFI data described in this paper are from the following two
sources:
\begin{enumerate}
\item  the ESO Large Programme 170.A-0789(A) (Principal In\-vest\-igator:
  J. Krautter, as chair of the SWG) which has accumulated data from
  January 27, 2003 to March 24, 2004 at the time of writing.

\item the contributing programs 70.A-0529(A); 71.A-0110(A);
  71.A-0110(B) with P. Schneider as the Principal In\-vest\-igator, which
  have contributed data from October 14, 2002 to September 29, 2003
\end{enumerate}

Observations were performed in the $B$-, $V$-, $R$-, and
$I$-passbands. These were split into OBs consisting of a sequence of
five (ten in the $I$-band) dithered sub-exposures with the typical
exposure time given in Table~\ref{tab:strategy}. The table gives: in
Col.1~ the passband; in Col.~2 the filter id adopting the unique
naming convention of the La Silla Science Operations Team; in Col.~3
the total exposure time in seconds; in Col.~4 the number of observing
blocks (OBs) per field; and in Col.~5 the integration time of the
individual sub-exposures in the OB. The dither pattern with a radius
of 80\arcsec\ was optimized for the best filling of the gaps. Filter
curves can be found in \citet{2001A&A...379..740A} and on the web page
of the La Silla Science Operations
Team.\footnote{\url{http://www.ls.eso.org/lasilla/sciops/2p2/E2p2M/WFI/filters/}}

Even though the nominal total survey exposure time for the $R$-band is
3500~s, the data contributed by the Bonn group provided additional
exposures totaling 11\,500~s each, spread over 4~OBs. For the
  same reason the $B$-band data for the field XMM-07 has a
  significantly larger exposure time than that given in
  Table~\ref{tab:strategy} (see Table~\ref{tab:img-products}).

\begin{table}
  \centering
  \caption{Planned observing strategy for the XMM-Newton follow-up
    survey.} 
  \begin{tabular}{llrrr}
    \hline\hline
    Passband & Filter & $T_\mathrm{tot}$ (s) & $N_\mathrm{OB}$ &
    $T_\mathrm{exp}$ (s) \\\hline 
    $B$ & B/123\_ESO879  & 1800 & 1 & 360 \\
    $V$ & V/89\_ESO843   & 4400 & 2 & 440 \\
    $R$ & Rc/162\_ESO844 & 3500 & 1 & 700 \\
    $I$ & I/203\_ESO879  & 9000 & 3 & 300 \\\hline
  \end{tabular}
  \label{tab:strategy}
\end{table}

Service mode observing provides the option for constraints on e.g.,
seeing, transparency, and airmass to be specified in order to meet the
requirements of the survey. The adopted constraints were: (1) dark sky
with a fractional lunar illumination of less than $0.4$; (2) clear sky
with no cirrus though not necessarily photometric; and (3) seeing
$\leq 1\farcs2$. The $R$-band images of the contributing program were
taken with a seeing constraint of $\lsim 1\farcs0$ so that the data
can be used for weak lensing studies.

The total integration time in some fields may be higher than the
nominal one listed in Table~\ref{tab:strategy} because unexpected
variations in ambient conditions during the execution of an OB can
cause, for instance, the seeing and transparency to exceed the
originally imposed constraints. If this happens, the OB is normally
executed again at a later time. In these cases the decision of using
or not all the available data must be taken during the data reduction
process. In the case of the present survey all available data were
included in the reduction, which explains why in some cases the total
integration time exceeds that originally planned.

This paper describes data accumulated prior to October 16, 2003,
amounting to about 80~h on-target integration. The science data
comprises 720 exposures split into 130 OBs. About 15\% of the
$B$-band and 85\% of the $R$-band data are from the contributing
programs.

\section{Data reduction}
\label{sec:reduction}
The accumulated optical exposures were reduced and calibrated using
the EIS Data Reduction System (da Costa et al., in preparation) and its
associated image processing engine based on the C++ EIS/MVM library
routines \citep[][Vandame et al., in
preparation]{2004PhDVandame}.\footnote{The PhD thesis is available
  from \url{http://www.eso.org/science/eis/publications.html}} This
library incorporates routines from the multi-resolution visual model
package (MVM) described in \citet{1995SigPr.....46..345R} and
\citet{1997ExA.....7..129R}. It was developed by the EIS project
to enable handling and reducing, using a single environment,
the different observing strategies and the variety of
single/multi-chip, optical/infrared cameras used by the different
surveys carried out by the EIS team. The platform independent EIS/MVM
image processing engine is publicly available and can be retrieved
from the EIS
web-pages.\footnote{\url{http://www.eso.org/science/eis/}}

The system automatically recognizes calibration and science exposures
and treats them accordingly. For the reduction, frames are associated
and grouped into \emph{Reduction Blocks} (RBs) based on the frame
type, spatial separation and time interval between consecutive
frames. The end point of the reduction of an RB is a
\emph{reduced image} and an associated weight map describing the local
variations of noise and exposure time in the reduced image. The data
reduction algorithms are fully described in
\citet{2004PhDVandame}. 

In order to produce a \emph{reduced image}, the individual exposures
within an RB are: (1) normalized to 1~s integration; (2)
astrometrically calibrated with the Guide Star Catalog version
  2.2 (GSC-2.2) as reference catalog, using a second-order polynomial
  distortion model; (3) warped into a user-defined reference grid
(pixel, projection and orientation), using a third-order Lanczos
kernel; and (4) co-added only using the weight for discarding the flux
contribution from masked pixels (e.g. satellite tracks
automatically detected and masked using a Hough transformation),
for which the pixel value is zero.  Note that individual exposures in
the RB are not scaled to the same flux level. This assumes that the
time interval corresponding to an RB is small enough to neglect
significant changes in airmass.

The 720 raw exposures were converted into 160 fully calibrated reduced
images, of which 146 were released in the $B$- (36), $V$- (32), $R$-
(43) and $I$- (35) passbands. Of the remaining 14, 10 were observed
with wrong coordinates, three (XMM-05 ($R$), XMM-06 ($I$), XMM-12
($V$)) were rejected after visual inspection and one (XMM-12) was
discarded due to a very short integration time (73~s), associated to a
failed OB. The number of reduced images (150) exceeds that of OBs
(130) because the RBs were built by splitting the OBs in order to
improve the cosmetic quality of the final stacked images, as discussed
below.

The photometric calibration of the reduced images was obtained using
the photometric pipeline integrated to the EIS data reduction system
as described in more detail in Appendix~\ref{sec:photometry}. In
particular, the XMM-Newton survey data presented here were obtained in
41 different nights of which 37 included observations of standard star
fields. For these 37 nights it was attempted to obtain photometric
solutions. The four nights without standard star observations are:
February 2, 3 and 4, 2003 (Public Survey); and November 8, 2002
(contributing program). For the nights with standard star
observations, the number of measurements ranges from a few to over
300, covering from 1 to 3 Landolt fields.

\begin{table}
\centering
  \caption{Summary of the number of nights with standard star observations
    and type of solution.}
  \begin{tabular}{lrrrrr}
    \hline\hline
    Passband &  default & 1-par & 2-par & 3-par & total \\
    \hline
    $B$ &  0 & 3 & 4 & 3 & 10 \\
    $V$ &  0 & 8 & 3 & 5 & 16 \\
    $R$ &  0 & 8 & 3 & 3 & 14 \\
    $I$ &  4 & 8 & 2 & 5 & 19 \\
    \hline
  \end{tabular}
  \label{tab:bestsol}
\end{table}

Table~\ref{tab:bestsol} summarizes the available photometric
observations. The table lists: in Col.~1 the passband; in Col.~2--5
the number of nights assigned a default solution or a 1--3-parameter
solution; and in Col.~6 the total number of nights with standard star
observations. For three nights (March 26, 2003; April 2, 2003; August
6, 2003) the solutions obtained in the passbands $V$, $I$, $R$,
respectively (either 2- or 3-parameter fits) deviate from the median by
$-0.26$, $-0.5$, $-0.25$ mag. Of those, only the $I$-band zeropoint
obtained for April 2, 2003 deviates by more than $3\sigma$ from the
solutions obtained for other nights. Note that the type of solution
obtained depends on the available airmass and color coverage, which in
the case of the XMM-Newton survey depends on the calibration plan adopted by
the La Silla Science Operations Team.

Because the EIS Survey System automatically carries out the
photometric calibrations it is interesting to compare the solutions to
those obtained by other means. Therefore, the automatically computed
3-parameter solutions of the EIS Survey System are compared with the
\emph{best solution} recently obtained by the La Silla Science
Operations Team. The results of this comparison are presented in 
Table~\ref{tab:photcomp} which lists: in Col.~1 the passband; in
Cols.~2--4 the mean offsets in zeropoint ($ZP$), extinction ($k$) and
color term (color), respectively. The agreement of the solutions is
excellent for all passbands. However, it is worth emphasizing that the
periods of observations of standard stars available to the two teams
do not coincide.

\begin{table}
  \centering
  \caption{Comparison between the EIS 3-parameter fit solutions and
    the Telescope Team's best solution.}
  \begin{tabular}{lrrr}
    \hline\hline
    Passband &  $\Delta ZP$ & $\Delta k$ & $\Delta$ color     \\
    \hline
    $B$ &  $0.00$ & $-0.03$ & $-0.06$  \\
    $V$ & $-0.05$ &  $0.00$ & $-0.01$  \\
    $R$ &  $0.00$ &  $0.10$ &  $0.00$  \\
    $I$ & $-0.04$ &  $0.05$ & $-0.02$  \\
    \hline
  \end{tabular}
  \label{tab:photcomp}
\end{table}

Not surprisingly, larger offsets are found when 2- and 1-parameter
fits are included, depending on the passband and estimator used to
derive the estimates for extinction and color term. Finally, taking
into consideration only 3-parameter fit solutions and after rejecting
$3\sigma$ outliers one finds that the scatter of the zeropoints is
$\lesssim 0.08$~mag. This number is still uncertain given the small
number of 3-parameter fits currently available, especially in the
$R$-band. The obtained scatter is a reasonable estimate for the current
accuracy of the absolute photometric calibration of the XMM-Newton survey
data.

There are two more points that should be considered in evaluating the
accuracy of the photometric calibration of the present data. First,
for detectors consisting of a mosaic of individual CCDs it is
important to estimate and correct for possible chip-to-chip variations
of the gain. For the present data these variations were estimated by
comparing the median background values of sub-regions bordering
adjacent CCDs. The determined variations were used to bring the gain
to a common value for all CCDs in the mosaic. This was applied to
both science and standard exposures. Second, it is also known that
large-scale variations due to non-uniform illumination over the field
of view of a wide-field camera exist. The significance of this effect
is passband-dependent and becomes more pronounced with increasing
distance from the optical axis
(\citealp{2001Msngr.104...16M,2004AN....325..299K}; Vandame et~al. in
preparation). Automated software to correct for this effect has been
developed but due to time constraints it has not yet been applied to
these data.

The final step of the data reduction process involves the assessment
of the quality of the reduced images. Following visual inspection,
each reduced image is graded, with the grades ranging from A (best) to
D (worst). This grade refers only to the visual aspect of the data
(e.g. background, cosmetics). Out of 150 reduced images covering (see
Sect.~\ref{sec:observations}) the selected XMM-Newton fields, 104 were
graded A, 35 B, 7 C and 4 D. The images with grades C and D are listed
in Table~\ref{tab:red-grade}. The table, ordered by field and date,
lists: in Col.~1 the field name; in Col.~2 the passband; in Col.~3 the
civil date when the night started (YYYY-MM-DD); in Col.~4 the grade
given by the visual inspection; and in Col.~5 the primary motive for
the grade. It is important to emphasize that the reduced images must
be graded, as grades are used in the preparation of the final image
stacks. In particular, reduced images with grade D have no
  scientific value and were not released and were discarded in the
stacking process discussed in the next section.

\begin{table*}
\centering
\caption{Grades representing the visual assessment of the reduced
  images.} 
\label{tab:red-grade}
\begin{tabular}{lcccl}
\hline\hline
Field  & Passband & Date & Grade & Comment \\\hline
XMM-05 & $R$ &  2002-10-14 & D & strong stray light contamination \\
XMM-06 & $I$ &  2003-01-29 & D & inadequate fringing correction \\
XMM-12 & $I$ &  2003-03-29 & D & very short integration time \\
XMM-12 & $V$ &  2003-09-27 & D & out-of-focus \\ 
XMM-01 & $V$ &  2003-02-01 & C & strong shape distortions \\
XMM-07 & $R$ &  2003-08-06 & C & stray light contamination \\
XMM-10 & $R$ &  2003-08-06 & C & fringing \\
XMM-10 & $R$ &  2003-09-23 & C & fringing \\
XMM-10 & $R$ &  2003-09-29 & C & fringing \\  
XMM-10 & $R$ &  2003-09-29 & C & fringing \\
XMM-10 & $R$ &  2003-09-29 & C & fringing \\
\hline
\end{tabular}
\end{table*}

The success rate of the automatic reduction process is better than
95\% and most of the lower grades are associated with observational
problems rather than inadequate performance of the software operating
in an un-supervised mode. An interesting point is that occasionally
$R$-band images are also affected by fringing (see Table
\ref{tab:red-grade}) -- for instance, in the
nights of August 6 and September 23 and 29, 2003, all from the
contributing program. The night of August 6 is one of the nights for
which the computed $R$-band zeropoint deviates from the median. This
points out the need to consider applying fringing correction also in
the $R$-band, at least in some cases. The $R$-band fringing problem
accounts for five out of seven grade C images. The remaining cases are
due to stray-light and strong shape distortions.

It should also be pointed out that the reduced images show a number of
cosmic ray hits. This is because the construction of RBs was optimized
for removing cosmic ray features in the final stacks using a
thresholding technique. To this end the number of images in an RB was
minimized for some field and filter combinations to have at least
three reduced images entering the SB. 

\section{Final products}
\label{sec:products}
\subsection{Images}
\label{sec:image-products}
The 146 reduced images with grades better than D were converted into
44 stacked (co-added) images using the EIS Data Reduction System. The
system creates both a final stack, by co-adding different reduced
images taken of the same field with the same filter (see
Appendix~\ref{sec:image-stacks}), and an associated product log with
additional information about the stacking process and the final image.
Note that all stacks (and catalogs) and their associated product logs
are publicly available from the EIS survey release and ESO
  Science Archive Facility
  pages.\footnote{\url{http://www.eso.org/science/eis/surveys/release_XMM.html}
  for catalogs and
    \url{http://archive.eso.org/archive/public_datasets.html} for the
    latest release of stacked images made in July 2005.}

\begin{figure*}[ht]
  \centering
  \includegraphics[width=17cm]{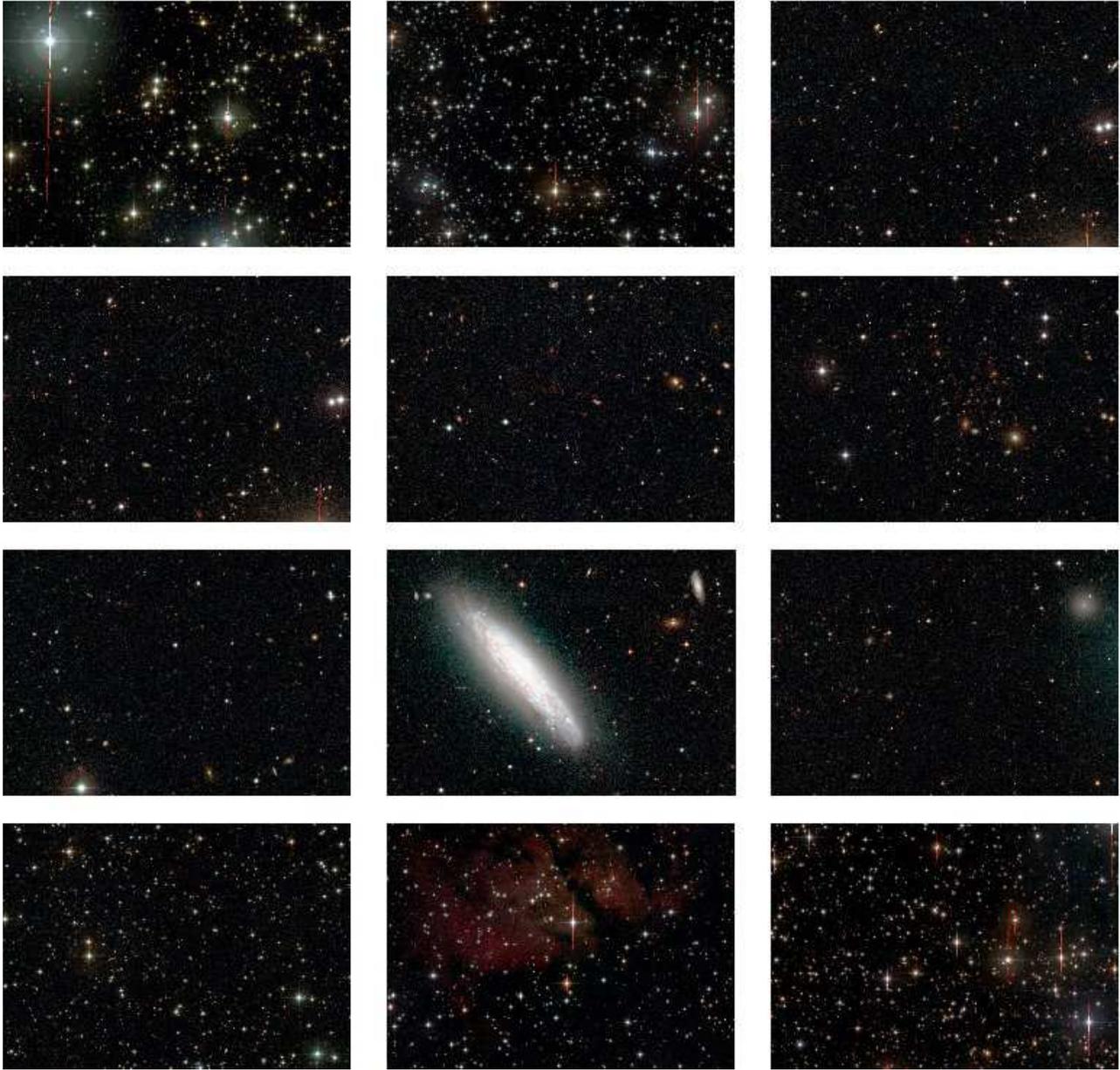}
  \caption{Above are cut-outs from color images of XMM-01 to XMM-12
    (from top left to bottom right) to illustrate the wide variety of
    fields the pipeline can successfully handle. The color images are
    $BVR$ composite were $R$-band data is available, $BVI$
    otherwise. The side length of the images displayed here is
    $7\farcm{9} \times 5\farcm{6}$. In these images North is up and
    East is to the left. These composite color images also demonstrate
    the accuracy of the astrometric calibration independently achieved
    in each passband.}
  \label{fig:xmm-overview}
\end{figure*}

The final stacks are illustrated in Fig.~\ref{fig:xmm-overview} which
shows cutouts from color composite images of the 12 fields. From this
figure, one can easily see the broad variety of fields observed by
this survey -- dense stellar fields (XMM-01, XMM-02, XMM-12),
sometimes with diffuse emission (XMM-11), extended objects (e.g.
XMM-08), and empty fields at high galactic latitude (e.g. XMM-07). While
the constraints imposed by the system normally lead to good results,
visual inspection of the images after stacking revealed that at least
in one case the final stacked image was significantly degraded by the
inclusion of a reduced image (graded B) with high-amplitude noise.
Therefore, this image was not included in the production of the
corresponding stack. The reason for this problem is being investigated
and may lead to the definition of additional constraints for the
automatic rejection algorithm being currently used.

Before being released the stacks were again examined by eye and
graded. Out of 44 stacks, 33 were graded A, 10 B, and 1 C, with no
grade D being assigned. In addition to the grade a comment may be
associated and a list of all images with some comment can be found in
the README file associated to this release in the EIS web-pages. The
comments refer mostly to images with poor background subtraction
either due to very bright stars (XMM-12) or extended, bright galaxies
(XMM-08, XMM-09) in the field. It is important to emphasize that the
reduction mode for these data was optimized for extragalactic,
non-crowded fields, which is not optimal for some of these fields.
Residual fringing is also observed in some stacks such as that of
XMM-10 in the $R$-band and XMM-04, XMM-06 in the $I$-band.

As mentioned in the previous section, to improve the rejection of
cosmic rays, the RBs were constructed so that in most cases the stack
blocks (SB) consist of at least 3 reduced images as input. This allows
for the use of a thresholding procedure, with the threshold set to
$2.5\sigma$, to remove cosmic ray hits from the final stacked
image. Even with this thresholding the stacks consisting of only three
RBs (totaling 5 exposures), mostly $B$-band images, still show some
cosmic ray hits. This happens primarily in the regions of the
inter-chip gaps, where fewer images contribute to the final
stack. Also, the automatic satellite track masking algorithm has
proven to be efficient in removing both bright and faint tracks. The
most extreme case is 3~satellite tracks of varying intensity in a
single exposure. The regions affected by satellite tracks in the
original images were flagged in the weight-map images and thus are
properly removed from the stacked image. Naturally, in the regions
where a satellite track was found in one of the contributing images
the noise is slightly higher in the stacked image. This is also
reflected in the final weight-map image.

The accuracy of the final photometric calibration of course depends on
the accuracy of the photometric calibration of the reduced images
which are used to produce the final co-added stacks and the number of
independent photometric nights in which these were observed (see
Table~\ref{tab:nights}). The former depends not only on the quality of
the night but also on the adopted calibration plan. To preview the
quality of the photometric calibration, Table~\ref{tab:nights}
provides information on the number of reduced images and number of
independent nights for each passband and filter. The table gives for
each field in: Col.~1 the field identification; Cols.~2--4 for each
passband the number of reduced images with the number in parenthesis
being the number of independent nights in which they were observed.
Complementing this information Table~\ref{tab:zp-field} shows the best
type of solution available for each field/filter combination. The
table gives: in Col.1 the field name; in Cols. 2--5 the number of free
parameters in the type of the \emph{best solution} available for the
passbands indicated. Solutions with more free parameters in general
indicate better airmass and color coverage, yielding better
photometric calibration. Examination of these two tables provide some
insight into the quality of the photometric calibration of each final
stack, as reported below.

\begin{table}
  \centering
  \caption{Summary of available data -- number of reduced images and
    in parentheses number of independent nights -- for each field and passband.}
  \label{tab:nights}
  \begin{tabular}{lrrrr}
    \hline\hline
    Field  &  $B$  & $V$  & $R$ & $I$  \\
    \hline
    XMM-01 & 3 (1) & 3 (2)  & 3 (1) & 3 (3)\\
    XMM-02 & 3 (1) & 3 (1)  & 3 (1) & 3 (1)\\ 
    XMM-03 & 3 (1) & 3 (2)  & 5 (3) & 3 (1)\\
    XMM-04 & 3 (1) & 3 (2)  & 4 (2) & 3 (2)\\
    XMM-05 & 3 (1) & 3 (2)  & 5 (1) & 3 (2)\\ 
    XMM-06 & 3 (1) & 3 (2)  & 6 (4) & 3 (2)\\
    XMM-07 & 3 (2) & 3 (2)  & 6 (4) & 3 (2)\\
    XMM-08 & 3 (1) & 3 (1)  & ---   & 3 (2)\\
    XMM-09 & 3 (1) & 3 (2)  & ---   & 3 (2)\\
    XMM-10 & 3 (1) &   ---  & 5 (3) & ---  \\
    XMM-11 & 3 (1) & 3 (2)  & 3 (1) & 5 (4)\\ 
    XMM-12 & 3 (1) & 2 (1)  & 3 (1) & 3 (2)\\
    \hline
  \end{tabular}
\end{table}

\begin{table}
\centering
\caption{Type of best photometric solution available for each field.}
\label{tab:zp-field}
\begin{tabular}{lcccc}
\hline\hline
Field & Default & 1-par & 2-par & 3-par \\\hline
 XMM-01 & $R$  & $BV$ & --- & $I$   \\
 XMM-02 & $RI$ & $BV$ & --- & ---   \\ 
 XMM-03 & ---  & $I$  & $R$ & $BV$  \\
 XMM-04 & ---  & $V$  & --- & $BRI$ \\
 XMM-05 & ---  & $R$  & $I$ & $BV$  \\
 XMM-06 & ---  & $V$  & $BR$& $I$   \\
 XMM-07 & ---  & $I$  & $B$ & $VR$  \\
 XMM-08 & ---  & $V$  & --- & $BI$  \\
 XMM-09 & ---  & ---  & $B$ & $VI$  \\
 XMM-10 & ---  & ---  & $B$ & $R$   \\
 XMM-11 & ---  & ---  & --- & $BVRI$\\  
 XMM-12 & ---  & $BR$ & $V$ & $I$   \\
\hline
\end{tabular}
\end{table}

The main properties of the stacks produced for each field and filter
are summarized in Table~\ref{tab:img-products}. The table gives: in
Col.~1 the field identifier; in Col.~2 the passband; in Col.~3 the
total integration time $T_\mathrm{int}$ in seconds, of the final stack; in
Col.~4 the number of contributing reduced images or RBs; in Col.~5 the
total number of science frames contributing to the final stack; in
Cols.~6 and 7 the seeing in arcseconds and the point-spread function
(PSF) anisotropy measured in the final stack; in Col.~8 the limiting
magnitude, $m_\mathrm{lim}$, estimated for the final image stack for a
2\arcsec\ aperture, $5\sigma$ detection limit in the AB system; in
Col.~9 the grade assigned to the final image during visual inspection
(ranging from A to D); in Col.~10 the fraction (in percentage) of
observing time relative to that originally planned.

\begin{table*}
  \centering
  \caption{Overview of the properties of the produced image
    stacks.}
  \begin{tabular}{llrrrrrrrr}
    \hline\hline
    Field & Passband & $T_\mathrm{int}$  & \#RBs & \#Exp. &
    Seeing & PSF rms & $m_\mathrm{lim}$  & Grade &
    Completeness\\ 
          &          & (s)&        &  & (arcsec) &  &
    (mag)  & &(\%) \\
\hline
XMM-01 & $B$ &  1800 & 3 &  5 & 1.19 & 0.056 & 24.94 & A & 100 \\
XMM-01 & $V$ &  6599 & 3 & 15 & 0.97 & 0.056 & 25.32 & A & 150 \\
XMM-01 & $R$ &  3500 & 3 &  5 & 0.82 & 0.074 & 23.97 & B & 100 \\
XMM-01 & $I$ &  8998 & 3 & 30 & 0.69 & 0.063 & 23.53 & A & 100 \\
\hline
XMM-02 & $B$ &  1800 & 3 &  5 & 1.17 & 0.051 & 24.51 & A & 100 \\
XMM-02 & $V$ &  4399 & 3 & 10 & 0.96 & 0.076 & 24.63 & A & 100 \\
XMM-02 & $R$ &  3500 & 3 &  5 & 0.64 & 0.087 & 24.69 & A & 100 \\
XMM-02 & $I$ &  5998 & 3 & 20 & 0.94 & 0.079 & 23.84 & A &  67 \\
\hline
XMM-03 & $B$ &  1800   & 3 &  5 & 1.01 & 0.031 & 25.44 & A & 100 \\
XMM-03 & $V$ &  4399   & 3 & 10 & 0.86 & 0.068 & 25.35 & A & 100 \\
XMM-03 & $R$ & 11\,748 & 5 & 20 & 0.83 & 0.152 & 25.15 & A & 336 \\
XMM-03 & $I$ &  9297   & 3 & 31 & 0.96 & 0.061 & 24.39 & A & 103 \\
\hline
XMM-04 & $B$ &  1800   & 3 &  5 & 1.17 & 0.041 & 25.22 & A & 100 \\
XMM-04 & $V$ &  4399   & 3 & 10 & 1.07 & 0.050 & 25.05 & A & 100 \\
XMM-04 & $R$ & 11\,748 & 4 & 20 & 0.76 & 0.069 & 25.57 & A & 336 \\
XMM-04 & $I$ &  8998   & 3 & 30 & 0.87 & 0.066 & 24.83 & A & 100 \\
\hline
XMM-05 & $B$ &  1800   & 3 &  5 & 1.24 & 0.076 & 25.18 & A & 100 \\
XMM-05 & $V$ &  4399   & 3 & 10 & 1.51 & 0.063 & 24.80 & A & 100 \\
XMM-05 & $R$ & 12\,348 & 5 & 21 & 0.94 & 0.072 & 25.58 & A & 353 \\
XMM-05 & $I$ &  8998   & 3 & 30 & 1.09 & 0.056 & 24.58 & A & 100 \\
\hline
XMM-06 & $B$ &  1800   & 3 &  5 & 0.87 & 0.052 & 25.57 & A & 100 \\
XMM-06 & $V$ &  4399   & 3 & 10 & 0.73 & 0.039 & 25.43 & A & 100 \\
XMM-06 & $R$ & 14\,998 & 6 & 25 & 0.85 & 0.060 & 24.54 & A & 429 \\
XMM-06 & $I$ &  8998   & 3 & 30 & 0.74 & 0.044 & 24.40 & A & 100 \\
\hline
XMM-07 & $B$ &  2699   & 2 &  8 & 1.24 & 0.035 & 25.55 & A & 150 \\
XMM-07 & $V$ &  4399   & 3 & 10 & 1.10 & 0.050 & 25.37 & A & 100 \\
XMM-07 & $R$ & 15\,698 & 6 & 27 & 1.03 & 0.058 & 25.66 & A & 449 \\
XMM-07 & $I$ &  8998   & 3 & 30 & 0.95 & 0.048 & 24.96 & A & 100 \\
\hline
XMM-08 & $B$ &  1800 & 3 &  5 & 1.28 & 0.062 & 25.62 & A & 100 \\
XMM-08 & $V$ &  4399 & 3 & 10 & 1.03 & 0.082 & 24.93 & A & 100 \\
XMM-08 & $I$ &  8998 & 3 & 30 & 0.79 & 0.052 & 24.76 & B & 100 \\
\hline
XMM-09 & $B$ & 1800 & 3 &  5 & 0.94 & 0.045 & 24.59 & B & 100 \\
XMM-09 & $V$ & 4839 & 3 & 11 & 0.83 & 0.031 & 24.20 & B & 110 \\
XMM-09 & $I$ & 8998 & 3 & 30 & 0.72 & 0.038 & 23.81 & B & 100 \\
\hline
XMM-10 & $B$ &  1500   & 3 &  5 & 1.12 & 0.042 & 24.26 & B &  83 \\
XMM-10 & $R$ & 11\,748 & 5 & 20 & 0.88 & 0.049 & 24.62 & C & 336 \\
\hline
XMM-11 & $B$ &  1800   & 3 &  5 & 1.09 & 0.058 & 25.25 & A & 100 \\
XMM-11 & $V$ &  4399   & 3 & 10 & 0.77 & 0.075 & 24.03 & A & 100 \\
XMM-11 & $R$ &  3500   & 3 &  5 & 0.60 & 0.090 & 23.10 & A & 100 \\
XMM-11 & $I$ & 12\,297 & 5 & 41 & 0.77 & 0.087 & 22.64 & A & 137 \\
\hline
XMM-12 & $B$ & 1800 & 3 &  5 & 1.09 & 0.087 & 23.41 & B & 100 \\
XMM-12 & $V$ & 3519 & 2 &  8 & 0.79 & 0.085 & 23.48 & B &  80 \\
XMM-12 & $R$ & 4899 & 3 &  7 & 0.64 & 0.111 & 23.16 & B & 140 \\
XMM-12 & $I$ & 3599 & 3 & 12 & 1.21 & 0.093 & 22.01 & B &  40 \\
\hline
  \end{tabular}
  \label{tab:img-products}
\end{table*}

This table shows that for most stacks the desired limiting magnitude
was met in $V$ (24.92~mag) or even slightly exceeded in $B$
(25.20~mag). The $R$- and $I$-band images are slightly shallower than
originally proposed with median limiting magnitudes of $24.66$~mag and
$24.39$~mag. Still, when only the high-galactic latitude fields are
included the median limiting magnitudes are fainter -- 25.33 ($B$),
25.05 ($V$), 25.36 ($R$) and 24.58 ($I$)~mag. All magnitudes are given
in the AB system. The median seeing of all stacked images is
$0\farcs94$ with the best and worst values being $0\farcs60$ and
$1\farcs51$, respectively. This is significantly better than the
seeing requirement of $1\farcs2$ specified for this survey.

Finally, the following remarks can be made concerning the image stacks
and their calibration:

\begin{itemize}
\item \textbf{XMM-01 ($R$)} -- The background subtraction near bright stars
  is poor. This field was observed as a single OB on February 3, 2003
  for which no standard stars were observed. Since this is a galactic
  field there are no complementary observations from the contributing
  program, and therefore these observations cannot be calibrated.

\item \textbf{XMM-01 ($I$)} -- This field at low galactic latitude is
  very crowded and no acceptable fringing map could be produced from the
  science exposures in the field. The de-fringing was done with an
  \emph{external fringing} map generated from science images taken on
  empty fields close in time to the XMM-01 $I$-band observations.

\item \textbf{XMM-02 ($R$)} -- The observations for this pointing and
  filter were done with one OB (5 exposures) on February 2,
  2003 for which no standard stars observations were carried out.
  
\item \textbf{XMM-02 ($I$)} -- The observations for this pointing and
  filter were done with two OBs (10 exposures each) on
  February 2, 2003 for which no standard stars observations were
  carried out. Like for XMM-01 ($I$) an external fringing map was
  used.

\item \textbf{XMM-03 ($V$)} -- The $V$-band calibration on the night of March
  26, 2003 yields a 3-parameter fit that deviates from the median of
  the solutions by roughly 0.26 mag (less than $3\sigma$).
  
\item \textbf{XMM-04 ($I$)} -- Low level fringing is still visible in
  the final stacked image.
  
\item \textbf{XMM-06 ($I$)} -- As in XMM-04, low level fringing is
  still visible in the final stack.
  
\item \textbf{XMM-07 ($B$)} -- From the three reduced images available
  only two were used for stacking because of the high amplitude of
  noise in one of them which greatly affected the final product.
  
\item \textbf{XMM-07 ($R$)} -- This field was observed in four nights
  (August 6, and September 23, 27, and 28, 2003) as part of the
  contributing program. For the night of August 6 a 3-parameter fit
  solution was obtained. However, this solution deviates by roughly
  0.25~mag relative to the median of all $R$-band solutions.
  
\item \textbf{XMM-07 ($I$)} -- There is a visible stray light
  reflection at the lower right corner of the image.
  
\item \textbf{XMM-08 ($V$)} -- The bright central galaxy is larger
  than the dithering pattern, thus making it difficult to estimate the
  background in its neighborhood. As a consequence the background
  subtraction procedure does not work properly.
  
\item \textbf{XMM-08 ($I$)} -- The comments about the background
  subtraction for the $V$-band image also apply to the $I$-band. This
  field was observed using 3 OBs (which in this case also correspond
  to 3 RBs) on two nights (March 30, 2003, one OB and April 2, 2003,
  two OBs). On the night of April 2 a 3-parameter solution was
  obtained for which the ZP determined deviates significantly (more
  than $3\sigma$) from the median of all solutions, even though the
  conditions of the night seem to have been adequate. The reason for
  this poor solution is at present unknown. Poor fringing correction
  is a possibility but needs to be confirmed. The zeropoint for the
  two reduced images taken in this night has been replaced by a
  default value.
  
\item \textbf{XMM-09 ($BVI$)} -- The preceding comment about
  background subtraction (see XMM-08) can be repeated here for the
  large galaxy in the North-West corner of the image. The background
  subtraction procedure fails, creating a strong variation around the
  galaxy.
  
\item \textbf{XMM-10 ($B$)}: This stack has a shorter exposure time
  than the others released, leading to higher background noise.
  
\item \textbf{XMM-10 ($R$)} -- This field was observed in the nights of
  August 6, and September 23 and 29, 2003 as part of contributing
  program. As in case of XMM-07 the solution for August 6
  deviates somewhat from the median.
  
\item \textbf{XMM-11 ($V$)} -- The same comments as for the photometric
  calibration of XMM-03 ($V$) apply to this image.

\item \textbf{XMM-11 ($I$)} -- Like for XMM-01 ($I$) an external
  fringing map was used. 

\item \textbf{XMM-12 ($BR$)} -- The background subtraction near
  bright stars is poor.
  
\item \textbf{XMM-12 ($V$)} -- The preceding comment about background
    subtraction also applies to this image. In addition the comment
    about the photometric calibration of XMM-03 ($V$-band) also applies
    to this image.
    
\item \textbf{XMM-12 ($I$)} -- The comment about background
    subtraction also applies to the $I$-band image. Like for XMM-01
    ($I$) an external fringing map was used.
\end{itemize}

Some improvements in the image quality may be possible in the future
by adopting a different observing strategy such as larger dithering
patterns to deal with more extended objects or shorter exposure times
to minimize the impact of fringing.

\subsection{Catalogs}
\label{sec:datacatalogs}

For the 8 fields located at high-galactic latitudes with $|b| >
30^\circ$, a total of 28 catalogs were produced (not all fields were
observed in all filters, see Table~\ref{tab:nights}). Catalogs for the
remaining low-galactic latitude fields were not produced since these
are crowded stellar fields for which SExtractor alone is not well
suited. As in the case of the Pre-FLAMES survey (Zaggia et al., in
preparation), it is preferable to use a PSF fitting algorithm such as
DAOPHOT \citep{1987PASP...99..191S}. Details about the catalog
production pipeline available in the EIS data reduction system are
presented in Appendix~\ref{sec:catalogs}.

As mentioned earlier, the fields considered here cover a range of
galactic latitudes of varying density of objects, in some cases with
bright point and extended sources in the field. In this sense this
survey is a useful benchmark to evaluate the performance of the
procedures adopted for the un-supervised extraction of sources and the
production of science-grade catalogs. This also required carrying out
tests to fine-tune the choice of input parameters to provide the best
possible compromise. Still, it should be emphasized that the catalogs
produced are in some sense general-purpose catalogs. Specific
science goals may require other choices of software (e.g. DAOPHOT,
IMCAT) and/or input parameters.

A key issue in the creation of catalogs is to minimize the number of
spurious detections and in general, the adopted extraction parameters
work well. However, there are unavoidable situations where this is not
the case. Among these are: (1) the presence of ghost images near
bright stars. Their location and size vary with position and magnitude
making it difficult to deal with them in an automatic way; (2) the
presence of bright galaxies because the algorithm for automatic
masking does not work well in this case; (3) residual fringing in the
image; (4) the presence of stray light, in particular, associated with
bright objects just outside the observed field; (5) when the image is
slightly rotated, the trimming procedure does not trim the corners of
the image correctly, leading to the inclusion of regions with a low
$S/N$. In these corners many spuriously detected objects are not
flagged as such. The XMM-Newton fields are a good showcase for these
various situations.

Another important issue to consider is the choice of the parameter
that controls the deblending of sources. Experience shows that the
effects of deblending depend on the type of field being considered
(e.g. empty or crowded fields, extended object, etc.) and vary across
the image. Some tests were carried out but further analysis of this
topic may be required.

A number of tests have also been carried out to find an adequate
compromise for the scaling factor used in the calculation of the size
of the automatic masks (see Appendix~\ref{sec:catalogs}) which depends
on the passband and the magnitude of the object. While the current
masking procedure generally works well, the optimal scaling will
require further investigation. It is also clear that for precision
work, such as e.g. lensing studies, additional masking by hand is
unavoidable. It should also be mentioned that occasionally the
masking of saturated stars fails. This occurs in five out of the 28
catalogs released and only for $\sim$10\% of the saturated stars in
them. These cases are likely to be of stars just barely saturated, at
the limit of the settings for automatic masking.

Bearing these points in mind, the following comments can be made
regarding some of the released catalogs:

\begin{itemize}
\item \textbf{XMM-03 ($B$)} -- The automatic masking misses a few
  saturated stars.
  
\item \textbf{XMM-06 ($B$)} -- Due to a small rotation of the image of a few
  degrees the trimming frame does not mask the borders completely.
  
\item \textbf{XMM-06 ($V$)} -- The deblending near bright
  galaxies is insufficient. Deblending near bright stars is too
  strong. 
  
\item \textbf{XMM-06 ($R$)} -- As in the $V$-band image the deblending
  near bright galaxies is insufficient.
  
\item \textbf{XMM-06 ($I$)} -- As in the $V$-band image the deblending
  near bright galaxies is insufficient. Spurious object detections are
  caused by reflection features of bright stars and stray light
  reflections.

\item \textbf{XMM-07 ($B$)} -- Spurious objects in the corners are
  caused by insufficient trimming. 
  
\item \textbf{XMM-08 ($B$)} -- Masks are missing for a number of
  saturated stars. XMM-08 contains an extended, bright galaxy (NGC
  4666) at the center of the image, plus a companion galaxy located
  South-East of it. The presence of these galaxies leads to a large
  number of spurious object detections in their surroundings in all
  bands.

\item \textbf{XMM-08 ($VRI$)} -- See the comments about spurious
  object detections for XMM-08 $B$-band.
  
\item \textbf{XMM-09 ($B$)} -- Cosmic rays are misidentified as real
  objects. The very bright galaxy located at the North-West of the
  image leads to the detection of a large number of spurious objects
  extending over a large area ($10\arcmin\times10\arcmin$) in all
  bands. Even though the galaxy has been automatically masked, the
  affected area is much larger than that predicted by the algorithm,
  which is optimized for stars. Thus, additional masking by hand would
  be required.
  
\item \textbf{XMM-09 ($VI$)} -- See the comments about spurious
  object detections for XMM-09 $B$-band.
  
\item \textbf{XMM-10 ($R$)} -- The stacked image was graded C because
  of fringing. The fringing pattern causes a high number of spurious
  object detections along the fringing pattern, leading to a catalog
  with no scientific value. \emph{This catalog is released exclusively
  as an illustration.}
\end{itemize}

\section{Discussion}
\label{sec:discussion}
\subsection {Comparison of counts and colors}
\label{sec:comp-counts-colors}
A key element in public surveys is to provide potential users with
information regarding the quality of the products released. To this
end a number of checks of the data are carried out and several
diagnostic plots summarizing the results are automatically produced
by the EIS Survey System. They are an integral part of the product
logs available from the survey release page. Due to the large number
of plots produced in the verification process these are not reproduced
here. Instead a small set illustrating the results are presented.

\begin{figure*}
  \sidecaption
  \includegraphics[width=12cm]{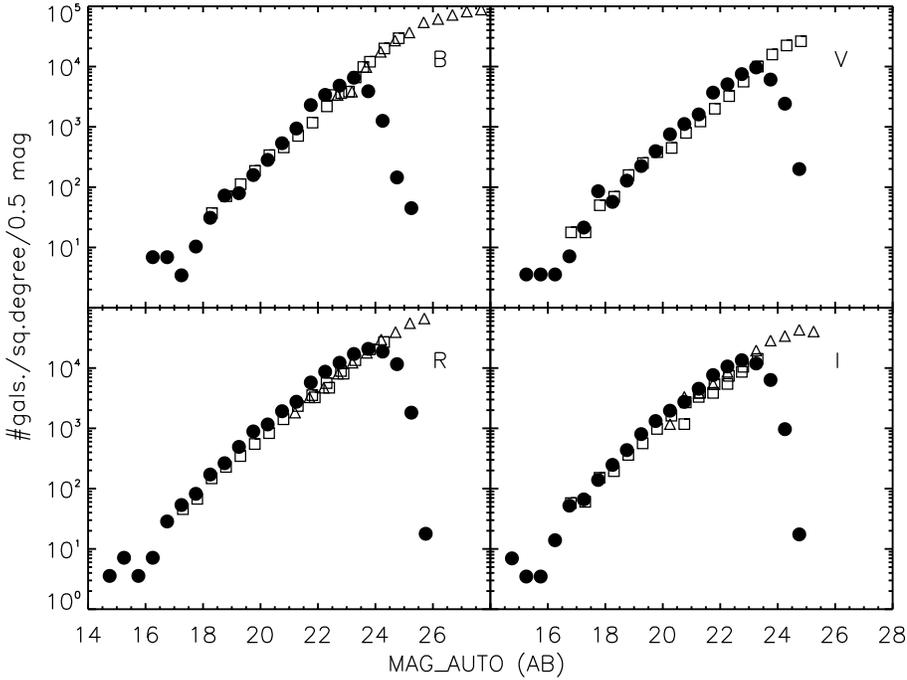}
  \caption{Galaxy number counts for the XMM-07 field for the different
    passbands as indicated in each panel. Full circles represent EIS
    data points, open triangles \citet{2001MNRAS.323..795M}, open
    squares~\citet{2001A&A...379..740A}.}
  \label{fig:xmm07counts}
\end{figure*}

A relatively simple statistics that can be used to check the catalogs
and the star/galaxy separation criteria is to compare the star and
galaxy number counts derived from the data to that of other authors
and/or to model predictions. As an example, Fig.~\ref{fig:xmm07counts}
shows the galaxy counts in different observed passbands for the field
XMM-07. Here objects with CLASS\_STAR$<0.95$ or fainter than the
object classification limit were used to create the sample of
galaxies. Note that the number counts shown in the figure take into
account the effective area of the catalog, which is available in its
\texttt{FIELDS} table (see Appendix~\ref{sec:catalogs}). As can be
seen, the computed counts are consistent with those obtained by
previous authors for all passbands \citep{2001A&A...379..740A,
  2001MNRAS.323..795M}.

A complementary test is to compare the stellar counts to those
predicted by models, such as the galactic model of \citet[][and
references therein]{2005A&A...436..895G}. Generally, the agreement of
model predictions is excellent for $B$- and $V$-band catalogs,
becoming gradually worse for $R$- and especially in $I$-band, near the
classification limit, with the counts falling below model predictions
(e.g. XMM-09 $I$-band). Note, however, that plots of CLASS\_STAR
versus magnitude show a less well defined stellar locus for these
bands. It is thus reasonable to assume that the observed differences
between catalogs and model predictions are due to misclassification of
stars as galaxies. Alternatively, these may also reflect short-comings
in the model adopted. However, a detailed discussion of this issue is
beyond the scope of the present paper.

\begin{figure}[t]
  \resizebox{\hsize}{!}{\includegraphics{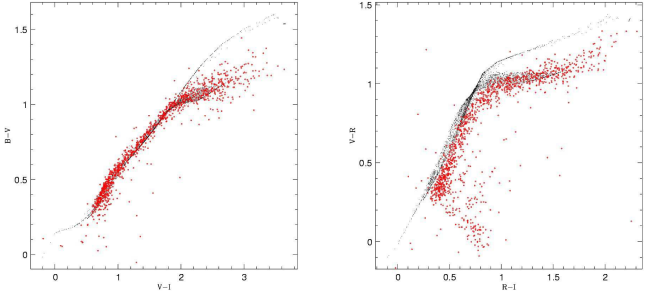}}
  \caption{$BVI$ (\emph{left panel}) and $VRI$ (\emph{right panel})
    color-color plot for stars objects in the XMM-06 field (large
    dots) and that obtained using theoretical models (small dots).}
  \label{fig:xmm06stellartracks}
\end{figure}

\begin{figure}[t]
  \resizebox{\hsize}{!}{\includegraphics{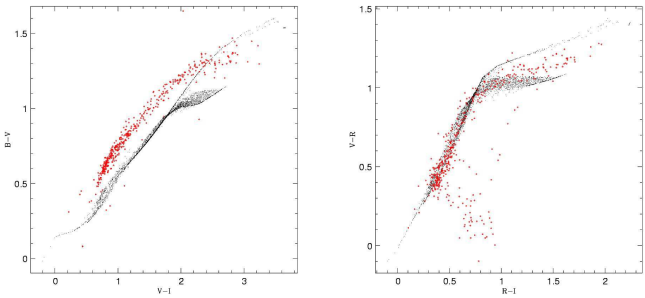}}
  \caption{Same as Fig.~\ref{fig:xmm06stellartracks} for XMM-07.}
  \label{fig:xmm07stellartracks}
\end{figure}

While useful to detect gross errors, number counts are not
sufficiently sensitive to identify more subtle differences. The
comparison of expected colors of stars with theoretical models
provides a better test of the accuracy of the photometric calibration
in the different bands. Using color transformations computed in the
same way as in \citet{2002A&A...391..195G}, the theoretical colors of
stars can be obtained. Such comparisons were made for all five fields
with data in four passbands. The results for two cases, XMM-06 and
XMM-07 are illustrated in Figs.~\ref{fig:xmm06stellartracks}
and~\ref{fig:xmm07stellartracks}, respectively, which show $(B-V)
\times (V-I)$ and $(V-R)\times (R-I)$ diagrams. For XMM-06 the data
are in excellent agreement with the colors of stars predicted by the
theoretical model, with only a small ($\lsim 0.05$ mag) offset in
$R-I$, indicating a good calibration. On the other hand, for XMM-07,
one observes a significant offset ($\sim 0.2$~mag) in $B-V$. This
field was chosen because it exemplifies the worst offset observed
relative to the theoretical models. Since this offset is only visible
in the $(B-V) \times (V-I)$ diagram, it suggests a problem in the
$B$-band data. Data for this field/filter combination comes from two
nights 2003-06-30 and 2003-08-06. Closer inspection of the
observations in the night of 2003-06-30 show that: (1) the
standard stars observations span only 2 hours in the middle of the
night; (2) the photometric zeropoint derived using the available
measurements (24.59~mag) is reasonably close ($\sim 1\sigma$) to the
median value of the long-term trend (24.71~mag); (3) the $B$
exposures were taken close to sunrise; and (4) there was a significant
increase in the amplitude of the DIMM seeing at the time the XMM-07
exposures under consideration were taken. For the night of 2003-08-06
standards cover a much larger time interval, yielding a
zeropoint of 24.82~mag with comparable difference relative to the
long-term median value for this filter as given above. The results
suggest that the observed problem is not related to the calibration of
the night, as will be seen below.

\subsection{Comparison with other reductions}
\label{sec:external-comparison}
\begin{table*}
  \centering
  \caption{Summary of the astrometric and photometric comparison. All
    differences were computed EIS$-$GaBoDS.}
  \begin{tabular}{lllrrr}
    \hline\hline
    Field & Target & Passband & $\Delta \alpha \cos(\delta)$ &
    $\Delta \delta$ & $\Delta m$ \\
         &         &          &  (arcsec) & (arcsec) & (mag) \\\hline

    XMM-03 & HE~1104$-$1805    & $B$ &  $0.02 \pm 0.04$ & $-0.01 \pm 0.03$ & $0.13
    \pm 0.02$ \\
    XMM-03 & HE~1104$-$1805    & $R$ &  $0.02 \pm 0.05$ & $-0.01 \pm 0.05$ & $0.01
    \pm 0.04$ \\
    XMM-04 & MS~1054.4$-$0321  & $B$ &  $0.00 \pm 0.05$ &  $0.00 \pm 0.05$ & $0.12
    \pm 0.04$ \\
    XMM-04 & MS~1054.4$-$0321  & $V$ &  $0.02 \pm 0.05$ & $-0.00 \pm 0.05$ & $0.00
    \pm 0.04$ \\
    XMM-04 & MS~1054.4$-$0321  & $R$ &  $0.03 \pm 0.06$ &  $0.00 \pm 0.06$ & $0.00
    \pm 0.03$ \\
    XMM-05 & BPM~16274         & $B$ &  $0.00 \pm 0.06$ & $-0.01 \pm 0.06$ & $0.05 \pm
    0.04$ \\
    XMM-05 & BPM~16274         & $R$ &  $0.03 \pm 0.09$ & $-0.01 \pm 0.09$ & $0.18 \pm
    0.04$ \\
    XMM-06 & RX J0505.3$-$2849 & $B$ &  $0.02 \pm 0.04$ &  $0.00 \pm 0.04$ &
    $0.00 \pm 0.02$ \\
    XMM-06 & RX J0505.3$-$2849 & $V$ &  $0.02 \pm 0.04$ & $-0.01 \pm 0.04$ &
    $-0.04 \pm 0.04$ \\
    XMM-06 & RX J0505.3$-$2849 & $R$ &  $0.01 \pm 0.03$ &  $0.01 \pm 0.04$ &
    $0.05 \pm 0.03$ \\
    XMM-07 & LBQS~2212$-$1759  & $B$ &  $0.02 \pm 0.06$ & $-0.02 \pm 0.06$ &
    $0.34 \pm 0.04$ \\
    XMM-08 & NGC~4666          & $B$ &  $0.00 \pm 0.06$ & $-0.01 \pm 0.05$ & $0.00 \pm
    0.03$ \\
    XMM-08 & NGC~4666          & $V$ &  $0.00 \pm 0.06$ & $-0.01 \pm 0.05$ & $-0.01 \pm
    0.03$ \\
    XMM-09 & QSO~B1246$-$057   & $B$ &  $0.02 \pm 0.06$ & $-0.01 \pm 0.05$ & $0.02
    \pm 0.02$ \\
    XMM-10 & PB~5062           & $B$ & $-0.03 \pm 0.05$ & $-0.01 \pm 0.07$ & $0.05 \pm
    0.02$ \\\hline 
  \end{tabular} 
  \label{tab:comparison}
\end{table*}
As shown above, comparison of different statistics, based on the
sources extracted from the final image stacks, to those of other
authors and to model predictions provide an internal means to assess
the quality of the data products. However, in the particular case of
this survey one can also benefit from the fact that about one third of
the accumulated data has been independently reduced by the Bonn group
in charge of the contributing program (Sect.~\ref{sec:observations}).
The images in common are used in this section to make a direct
comparison of the astrometric and photometric calibrations. In their
reduction, the Bonn group used their ``Garching-Bonn Deep Survey''
(GaBoDS) pipeline \citep{2005AN....326..432E}.

A total of 15 stacked images in the $B$-, $V$-, and $R$-bands were
produced and compared to those produced by the EIS/MVM pipeline.  The
astrometric calibration was done using the GSC-2.2 catalog, the same
as that of the EIS reduction. In contrast to the reductions carried
out by the EIS system, images were photometrically calibrated using
the measurements of standard stars compiled by
\citet{2000PASP..112..925S}. The type of solution (number of free
parameters of a linear fit) for a night was decided on a case by case
basis after visual inspection of the linear fits.

To carry out the comparison of the data products, catalogs were
produced from the EIS and GaBoDS images using the same extraction
parameters. These catalogs were associated with each other to produce
a merged catalog for each field and passband. The results of this
comparison for all the available images in common are presented in
Table~\ref{tab:comparison}. The table gives: in Col.~1 the field name;
in Col.~2 the original target name; in Cols.~3 and 4 the mean offset
and standard deviation in right ascension and declination in
arcseconds; in Col.~5 the mean and standard deviation of the magnitude
differences as measured within an aperture of 3\arcsec. The mean and
standard deviation of the magnitude differences were determined in the
interval $17<m<21$. This range was chosen to avoid saturated objects
at the bright end and to limit the comparison to objects whose
estimated error in magnitude is smaller than about $0.01$~mag at the
faint end. An iterative $5\sigma$ rejection, which allowed rejected
points to re-enter if they are compatible with later determinations of
the mean and variance, was employed to ignore obvious outliers in the
computation of the mean and the standard deviation.

\begin{figure}[ht]
  \resizebox{\hsize}{!}{\includegraphics{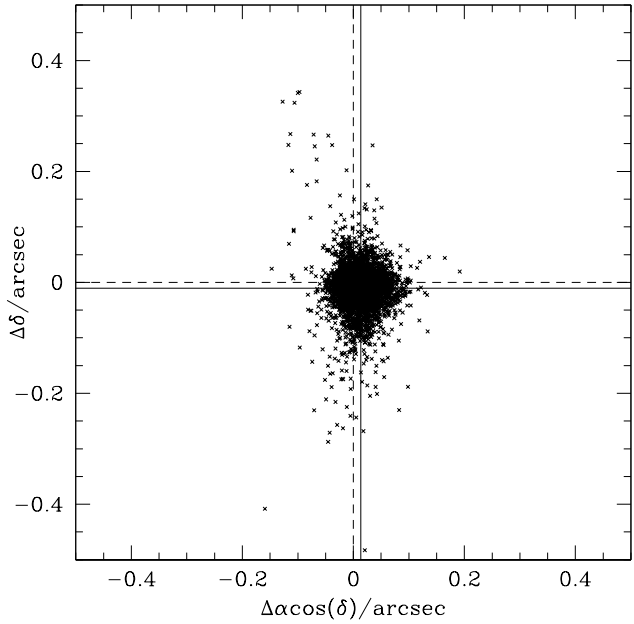}}
  \caption{Comparison of astrometry for the $R$-band image of
     XMM-06 (RX~J0505.3$-$2849), selected to represent a
    typical case. The offsets are
    computed EIS$-$GaBoDS. The dashed lines are centered on $(0,0)$,
    while the solid lines denote the actual barycenter of the points.}
  \label{fig:sharc2r.astrom}
\end{figure}

Fig.~\ref{fig:sharc2r.astrom} illustrates the results obtained from
the comparison of the position of sources extracted from images
produced by the two pipelines for the particular case of XMM-06 in
$R$-band. From the figure one can see that the positions of the
sources agree remarkably well. In fact, as summarized in
Table~\ref{tab:comparison}, the typical mean deviation is $\sim
20$~mas with a standard deviation of $\sim 50$~mas, confirming the
excellent agreement in the \emph{external} (absolute) astrometric
calibration to be distinguished from the \emph{internal} calibration
discussed later.

\begin{figure}[ht]
  \resizebox{\hsize}{!}{\includegraphics{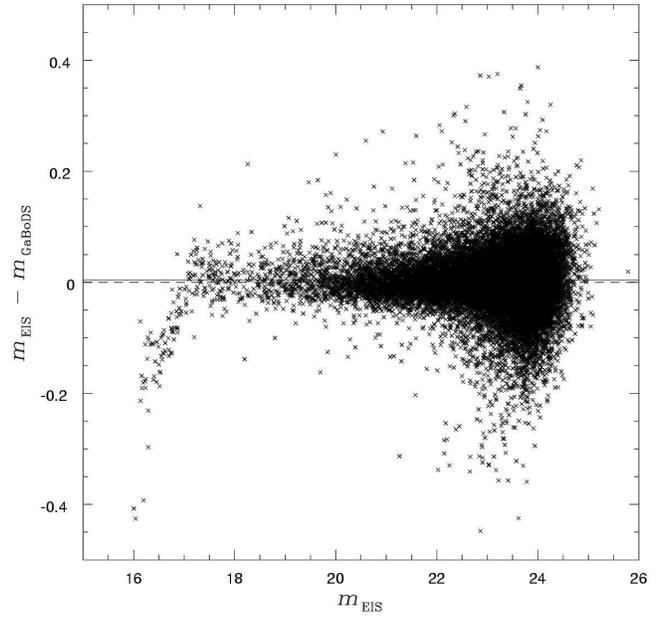}}
  \caption{Comparison of aperture magnitudes (3\arcsec\ aperture)
  measured on the $R$-band image of the field XMM-04
  (MS~1054.4$-$0321). The dashed line is at a magnitude difference of
  0, while the solid line denotes the actual offset between the EIS
  and the GaBoDS reduction. The difference at the bright end is caused
  by different treatments of saturated objects in both pipelines.}
  \label{fig:MS1054.R.EIS-GaBoDS}
\end{figure}

Fig.~\ref{fig:MS1054.R.EIS-GaBoDS} shows a plot of the magnitude
differences measured on the GaBoDS $R$-band image of the field
MS~1054.4$-$0321 (XMM-04) versus the magnitudes measured on the
corresponding EIS image. This field shows that the photometry of both
reductions agree remarkably well. The measured scatter of the
magnitude differences is small ($\sim 0.03$~mag) for this as well as
for most other fields. This result indicates that the internal
procedures used by the two pipelines to estimate chip-to-chip
variations are consistent. Moreover, inspection of the last column of
Table~\ref{tab:comparison} shows that for 11 out of 15 cases the mean
offsets are $\lsim$0.05~mag. This is reassuring for both pipelines
considering all the differences involved in the process, which include
differences in the routines, procedures and the standard stars
used. It is important to emphasize that differences in the computed
zeropoint of the photometric solutions are $\lsim$ 0.08~mag, even for
the cases with the largest differences such as XMM-05 ($R$) and XMM-07
($B$). The value of 0.08~mag is consistent with the scatter measured
from the long-term trend shown by the zeropoints computed over a large
time interval, as presented in the EIS release of WFI photometric
solutions, thus representing the uncertainty in the photometric
calibration. Therefore, the offsets reported in the table cannot be
explained by differences in the photometric calibration alone. This
point is investigated in more detail for XMM-05 $R$-band and XMM-07
$B$-band.

All $R$-band images for XMM-05 were taken in one night and the
photometric solutions determined by both teams agree very well. While
the source of the discrepancy has not yet been identified, the stellar
locus in the $(B-V) \times (V-I)$ and $(V-R) \times (R-I)$ diagrams
based on the source catalog extracted from the EIS images yield
results which are consistent with model predictions, suggesting that
the problem may lie in the Bonn reductions. On the other hand, the
large offset (0.34~mag) between the $B$-band observations of the field
XMM-07 is most likely caused by the data taken in the night
2003-06-30. While the standard star observations in this night suggest
relatively good photometric conditions, the available measurements
span only about 2 hours in the middle of the night, while the science
exposures were taken at the very end of the night. Inspection of the
ambient condition shows a rapid increase in the amplitude of the DIMM
seeing which could be related to a localized variation in the
transparency. In fact, the Bonn pipeline, which monitors the relative
differences in magnitude for objects extracted from different
exposures in an OB, finds strong flux variations that could be caused
by changes in the sky transparency or by the twilight at sunrise. The
latter could also account for the fact that these observations were
later repeated in August of that year. The important point is that the
Bonn group discarded the calibration of the frames taken in
2003-06-30, while the automatic procedure adopted by EIS did not.

\begin{figure*}
  \sidecaption
  \includegraphics[width=12cm]{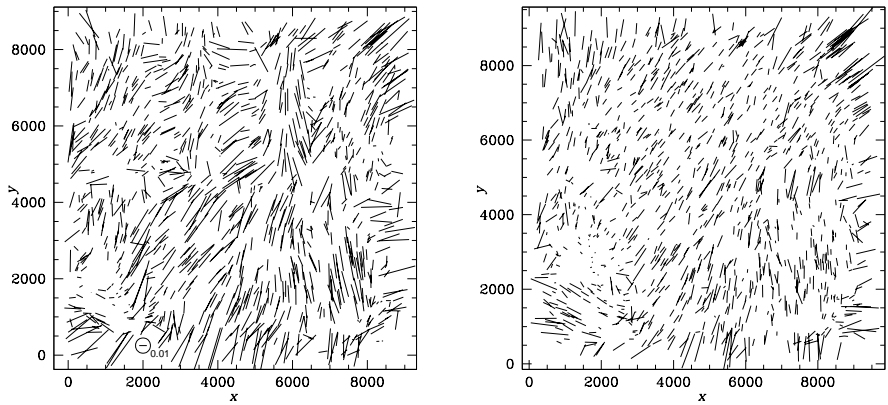}
  \caption{Stellar PSF pattern in the $R$-band images of the field
  MS~1054.4$-$0321 in the EIS reduction (left panel) and the GaBoDS
  reduction (right panel). The encircled stick in the left panel
  denotes an ellipticity of $\varepsilon = 0.01$. Both plots have the
  same scale.}
  \label{fig:psf_comp}
\end{figure*}

In addition to evaluating the accuracy of the image registration and
photometric calibration, the independent reductions also offer the
possibility to evaluate the shape of the images. To this end the PSF
of bright, non-saturated stars on the $R$-band images for XMM-04
(MS~1054.4$-$0321) and XMM-06 (RX~J0505.3$-$2849) were measured and
compared. These are the only two cases in which the final stacked
images were produced by using exactly the same reduced images. This
is caused by the differences in the criteria adopted in building the
SBs. In the case of XMM-06, one finds that the size and pattern of
the PSF are in good agreement and both reductions yield a smooth PSF
with no obvious effects of chip boundaries over the whole field. The
situation is different for XMM-04 as can be seen in
Fig.~\ref{fig:psf_comp}, which shows a map of the PSF distortion
obtained by the EIS (left panel) and Bonn (right panel) groups. While
the overall pattern of distortion is similar, the amplitude of the PSF
distortion of the EIS reduction is significantly larger and exhibits
jumps across chip borders. Although the effect is small in absolute
terms, it should be taken into account for applications relying on
accurate shape measurements. The reason for these differences is
likely due to the fact that the astrometric calibration in the EIS
pipeline is done for each chip relative to an absolute \emph{external}
reference, without using the additional constraint that the chips are
rigidly mounted to form a mosaic. By neglecting this constraint, the
solution for each chip in the mosaic may vary slightly depending on
the dithered exposure being considered and the density and spatial
distribution of the reference stars in and around the field of
interest. Since the accuracy of the GSC-2.2 of $250$~mas is
approximately equal to the pixel size of WFI of $0\farcs{238}$, in
addition to the absolute calibration of the image centroid, finding an
\emph{internal} relative astrometric solution further ensures that
images in different dithered exposures map more precisely onto each
other during co-addition. Imperfections in the \emph{internal}
relative astrometry result in objects not being matched exactly onto
each other, thereby degrading the PSF of the co-added image.

\subsection{X-ray/optical correlation}
\label{sec:x-rayopt-corr}
As pointed out in the introduction, the ultimate goal of this optical
survey has been to provide catalogs from which one can identify and
characterize the optical properties of X-ray sources detected with
deep XMM-Newton exposures.

X-ray source lists for the high-galactic latitude fields were produced
by the AIP-node of the SSC. These are based on
pipeline processed event lists which were obtained with the latest
official version of the Software Analysis System (SAS-V6.1). In its
current version this SAS-based pipeline does not work with stacked
images.

Source detection was performed as a three-stage process using
\texttt{eboxdetect} in local and in map mode followed by a multi-PSF
fit with \texttt{emldetect} for all sources present in the initial
source lists. The multi-PSF fit invoked here works on 15 input images,
i.e.~5 per EPIC camera. The five energy bands used per camera cover
the ranges: (1) 0.1--0.5~keV; (2) 0.5--1.0~keV; (3) 1.0--2.0~keV; (4)
2.0--4.5~keV; and (5) 4.5--12.0~keV.

The SAS task \texttt{eposcorr} was applied to the X-ray source list.
\texttt{Eposcorr} correlates the X-ray source positions with the
positions from an optical source catalog, in this case the EIS
catalog, to correct the X-ray positions, assuming that the true
counterparts are contained in the reference catalogue.

The source detection scheme used here is very similar to the pipeline
implemented for the production of the second XMM-Newton catalog of
X-ray sources to be published by the XMM-Newton-SSC later in 2005
(Watson et al., in preparation). This approach is superior to that
used for the creation of source lists which are currently stored in
the XMM-Newton Science Archive since it makes use of X-ray photons
from all cameras simultaneously. It also distinguishes between
point-like and extended X-ray sources. In this paper we only consider
point-like sources. Extended sources at high galactic latitudes are
almost exclusively galaxy clusters and cannot be matched with
individual objects in the optical catalogs. Examining their properties
is beyond the scope of this work.

In carrying out the matching between the XMM-Newton source lists and those
extracted from the optical images, it is important to note that the
X-ray images lie fully within the FOV of the WFI images. Hence an
optical counterpart can be potentially found for any of the X-ray
sources.

From the high-galactic latitudes there are 3 fields with more than one
observation. However, for XMM-05 only the two available observations
with good time $t > 10$~ks were considered. One of them had technical
problems that prevented it from being used for catalog extraction.
For XMM-09 the source list created contained many spurious sources due
to remaining calibration uncertainties in the pipeline processed
images and was not considered further. Figures showing the results of
the source detection process with all sources indicated on an image in
TIFF format, the composite X-ray images, and the source lists can be
found on the web-page of the
AIP-SSC-node.\footnote{\url{http://www.aip.de/groups/xray/XMM_EIS/}}
Below these source lists are used to identify their optical
counterparts.

The extraction yields 995 point-like X-ray sources of which 742 are
unique. The difference between these two numbers reflects differences
in the three independent source lists extracted from the field XMM-07.
The mean flux of the 742 unique X-ray sources is
$F_\mathrm{mean}(0.5-2.0\,\mathrm{keV}) = 8.5 \times
10^{-15}$\,erg\,cm$^{-2}$\,s$^{-1}$, the median flux in this band is
$F_\mathrm{med}(0.5-2.0\,\mathrm{keV}) = 3.7 \times
10^{-15}$\,erg\,cm$^{-2}$\,s$^{-1}$. Sources with
$F(0.5-2.0\,\mathrm{keV}) = 4 \times
10^{-15}$\,erg\,cm$^{-2}$\,s$^{-1}$ are detected already with an
exposure time of 5~ks, while the limiting flux in the EIS-XMM fields at the
deepest exposure levels is $F_\mathrm{lim} \simeq 3 \times
10^{-16}$\,erg\,cm$^{-2}$\,s$^{-1}$.

\begin{table*}
  \centering
  \caption{Contents of X-ray source lists for high-latitude XMM-EIS fields.}
  \begin{tabular}{llrcrrrrrrrrrrr}
    \hline\hline
    Field  & Obs. ID    & $N_\mathrm{s}$ & Passband & $N_\mathrm{m}$ &
    $N_\mathrm{1}$ & $(N_\mathrm{1}/N_\mathrm{s})(\%)$ &
    $N_\mathrm{all}$ & $N_\mathrm{1,all}$ & $B$ & $V$ & $I$ & $BV$ &
    $BI$ & $BVI$\\
    \multicolumn{1}{c}{(1)} & \multicolumn{1}{c}{(2)} & \multicolumn{1}{c}{(3)} & \multicolumn{1}{c}{(4)} & \multicolumn{1}{c}{(5)} & \multicolumn{1}{c}{(6)} & \multicolumn{1}{c}{(7)} & \multicolumn{1}{c}{(8)} & \multicolumn{1}{c}{(9)} & \multicolumn{1}{c}{(10)} & \multicolumn{1}{c}{(11)}
    & \multicolumn{1}{c}{(12)} & \multicolumn{1}{c}{(13)} &
    \multicolumn{1}{c}{(14)} & \multicolumn{1}{c}{(15)} \\\hline 
    XMM-03 & 0112630101 &  69 & $R$ &  92 &  61 &  88 &  62 &  47 & 
    1 & 1 & 1 &  0 &  0 &  0 \\
           &            &  69 &     &  36 &  35 &  51 &  27 &
    27 & 0 & 0 & 0 & 0 &  0 &  0 \\
    XMM-04 & 0094800101 & 101 & $R$ & 156 &  91 &  90 &  84 &  68 &  
    0 & 0 & 1 & 1 & 0 & 0 \\ 
           &            & 101 &     &  74 &  73 &  72 &  57 &  57 & 
    0 & 1 & 1 & 0 & 0 &  1 \\
    XMM-05 & 0125320401 &  89 & $R$ & 130 &  79 &  89 &  52 &  42 & 
    1 & 0 & 2 & 0 & 0 &  0 \\
           &            &  89 &     &  59 &  57 &  64 &  37 &  37 &
    1 & 0 & 2 & 1 & 0 & 0 \\
    XMM-06 & 0111160201 & 110 & $R$ & 173 & 101 &  92 & 105 &  79 &
    1 & 1 & 0 & 0 & 0 & 0 \\
           &            & 110 &     &  76 &  72 &  65 &  61 &  58 & 
    3 & 0 & 1 & 0 & 0 & 0 \\
    XMM-07 & 0106660101 & 144 & $R$ & 191 & 119 &  83 &  84 &  66 &
    3 & 0 & 2 & 1 & 0 & 2 \\
           &            & 144 &     &  82 &  81 &  56 &  52 &  52 &
    2 & 2 & 1 & 1& 0 & 1 \\
    XMM-07 & 0106660201 & 110 & $R$ & 134 &  91 &  83 &  65 &  53 &
    3 & 0 & 2 & 1 & 0 & 2 \\
           &            & 110 &     &  62 &  61 &  56 &  39 &  39 &
    1 & 0 & 0 & 2 & 1 & 0 \\
    XMM-07 & 0106660601 & 162 & $R$ & 211 & 139 &  86 &  93 &  76 &
    1 & 0 & 1 & 2 & 0 & 0 \\
           &            & 162 &     & 100 &  98 &  60 &  59 &  59 &
    1 & 1 & 1 & 2 & 0 & 1 \\
    XMM-08 & 0110980201 & 123 & $I$ & 130 &  97 &  79 &  82 &  70 &
    1 & 0 & --- & 0 & --- & --- \\
           &            & 123 &     &  73 &  73 &  59 &  58 &  58 &
    0 & 2 & --- & 1 & --- & --- \\
    XMM-10 & 0012440301 &  88 & $R$ & 113 &  76 &  86 &  50 &  46 &
    1 & --- & --- & --- & --- & --- \\
           &            &  88 &     &  57 &  56 &  64 &  35 &  34 &
    3 & --- & --- & --- & --- & --- \\\hline
  \end{tabular}
  \label{tab:xs}
\end{table*}

Nearly all the X-ray source lists were matched to catalogs extracted
from the $R$-band images, with the exception of field XMM-08, which
was correlated with the $I$-band. Two search radii, 2\arcsec\ and
5\arcsec, were used. The larger value reflects the typical statistical
error in X-ray source position determination (typically in the range
$\sim0.5\arcsec\text{--}2\arcsec$), coupled with an additional
systematic error component ($\sim1\arcsec$) in the attitude of the
spacecraft. Hence, a matching radius of 5\arcsec\ corresponds to
roughly a $2\text{--}3\sigma$ uncertainty for most of the sources.
The smaller correlation radius is justified by the distribution of the
positional accuracy of the X-ray sources, which peaks at $\sim
1\farcs3$. It extends up to 3\arcsec\ with the majority of sources
(92\%) being within 2\arcsec.

The results of X-ray source extraction and their cross-identification
with their optical counterparts for 7 high-galactic latitude fields (9
observations) are summarized in Table~\ref{tab:xs}. For each field two
rows are given: the first row refers to the matching done with a
5\arcsec\ search radius, in the second row the numbers for the smaller
2\arcsec\ search radius are reported. The table lists: in Col.~1 the
field name; in Col.~2 the Obs.~ID of the XMM-Newton observation; in
Col.~3 the number of detected X-ray point sources with a likelihood of
existence larger than \texttt{detml = 6}, $N_\mathrm{s}$; in Col.~4
the passband of the catalog used as the optical reference for
matching; in Col.~5 the number of matches, $N_\mathrm{m}$ within
5\arcsec\ (2\arcsec); in Col.~6 the number of X-ray sources with at
least one match $N_\mathrm{1}$. In the case of multiple matches the
$m=1$ sources refer to the closest matching optical source; in Col.~7
the identification rate $N_\mathrm{1}/N_\mathrm{s}$; in Col.~8, the
number of X-ray sources, which have at least one counterpart in the
optical reference catalog, and are also detected in all other
available optical passbands. This is a subset of the objects listed in
Col.~5; in Col.~9, the same as in the previous column but for the
$m=1$ optical counterparts, $N_\mathrm{1,all}$, which is a subset of
the objects listed in Col.~6; finally, in Col.~10--15 the number of
optical counterparts in other passbands, which are not detected in the
reference catalog.  In Col.~13--15 $BV$, $BI$ and $BVI$ refer to
objects which are simultaneously detected in the respective passband
but do not correspond to matches of X-ray sources with the reference
catalog.  Because we only list objects \emph{without} match in the
reference catalog the number of objects reported in Col.~10--15 is in
some cases higher in the second row than in the first row. These are
X-ray sources with matches in $B$-, $V$-, or $I$-band within a circle
of 2\arcsec\ having matches in the reference catalog only in the
larger 5\arcsec\ search radius.

\begin{figure*}
  \centering
  \includegraphics[angle=-90, width=17cm]{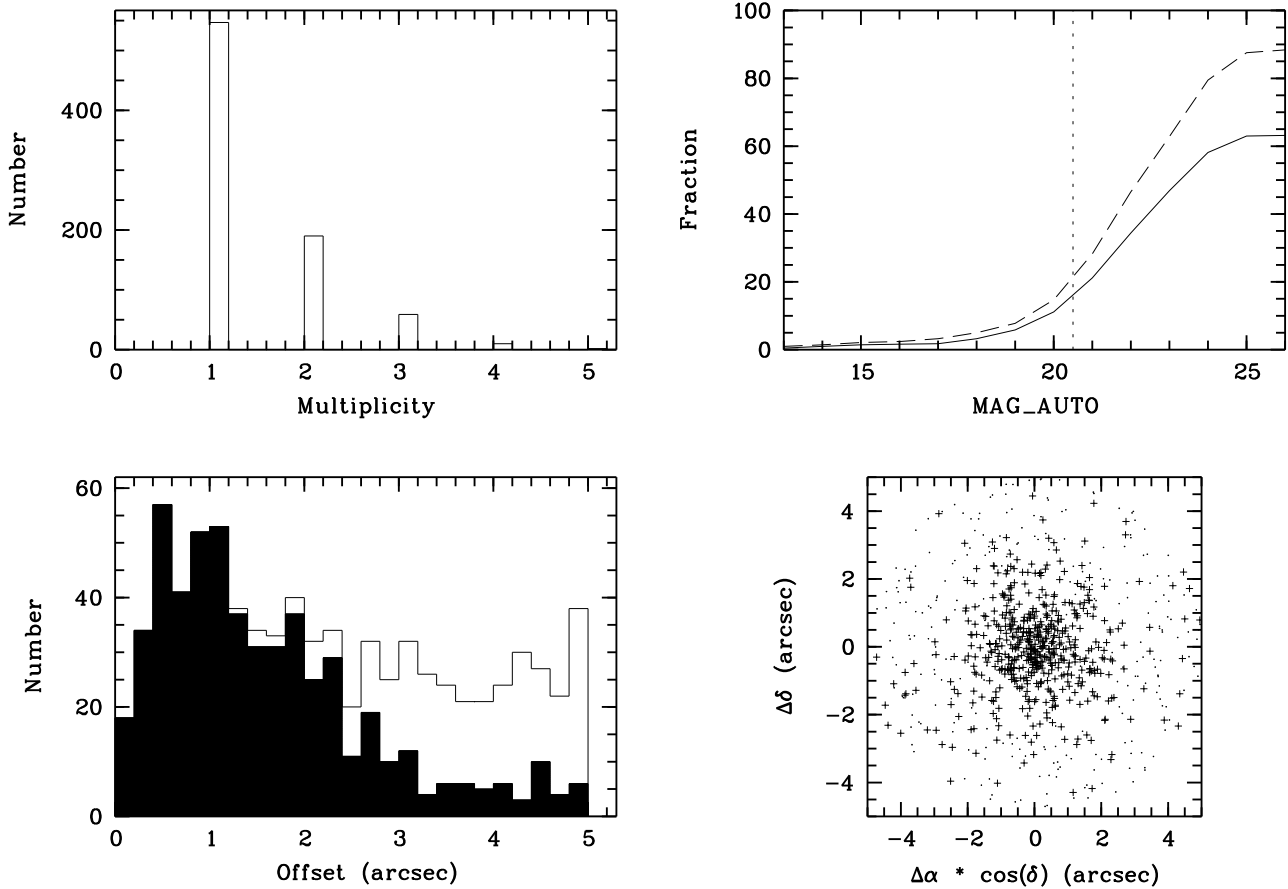}
  \caption{X-ray/optical $R$-band positional correlation. (\textit{top
      left}) number of correlated optical sources to X-ray sources
    within 5\arcsec; (\textit{top right}) cumulative fraction of X-ray
    sources with optical counterparts in the $R$-band catalog. The
    dashed line is for the 5\arcsec search radius, the straight line
    for the 2\arcsec search radius. The vertical short-dashed line
    denotes the approximate limit of the DSS; (\textit{bottom left})
    distribution of X-ray minus optical positional offset of all and
    $m=1$ sources; (\textit{bottom right}) distribution of positional
    offsets in the right ascension -- declination plane.}
\label{fig:xoc}
\end{figure*}

The results of the X-ray/optical cross-correlation for all fields with
available $R$-band catalogs (619 unique sources) are displayed in
Fig.~\ref{fig:xoc}. The figure shows: (top left) the multiplicity
function; (top right) the cumulative fraction of X-ray sources with
optical counterparts in a 5\arcsec search radius (dashed line) and a
2\arcsec search radius (straight line); (bottom left) the distribution
of the positional offsets between X-ray and optical sources; and
(bottom right) the corresponding scatter plot in the
$\alpha~\times~\delta$ plane.  Note that in three panels all $m=1$
matches are represented by filled histograms and/or larger symbols.

Inspection of Fig.~\ref{fig:xoc} shows that: (1) about 87\% (61\%) of
the X-ray point sources have at least one optical counterpart within
the search radius of 5\arcsec\ (2\arcsec) down to $R\sim25$~mag, and
very few sources have more than 3 matches. In only very few cases one
finds up to five associated optical sources, i.e.~potential physical
counterparts; (2) only about 15\% of the X-ray sources have
counterparts down to the Digital Sky Survey magnitude limit ($R \sim
20.5$), underscoring the need for dedicated optical imaging in order
to identify the X-ray source population; (3) the distribution of the
X-ray/optical positional offset peaks at around 1\arcsec\ for the
sources with $m=1$. The $m=1$ matches are well concentrated within a
circle of 2\arcsec. The distribution is almost flat if all
associations are considered. This underlines that the true physical
counterparts to the X-ray sources will be found predominantly among
the $m = 1$ sources, i.e.~the nearest and in most cases single
associated optical sources; (4) the positional differences between
X-ray and optical coordinates seem to be randomly distributed.

The statement that the additional correlations found within the
greater search radius are chance alignments is strengthened by an
estimate of the number of random matches between X-ray and optical
sources. Using an average of 110 X-ray sources per field we can
compute the total area covered by the search circles with 5\arcsec
(2\arcsec) radius.  Multiplying this with the typical number density
of sources in the optical catalogs (30~arcmin$^{-2}$) we estimate 70
(12) random matches for an average field. The number of random matches
within the smaller correlation radius is well below the observed number
of optical/X-ray counterparts.

\begin{figure*}
  \centering
  \includegraphics[angle=-90,width=17cm]{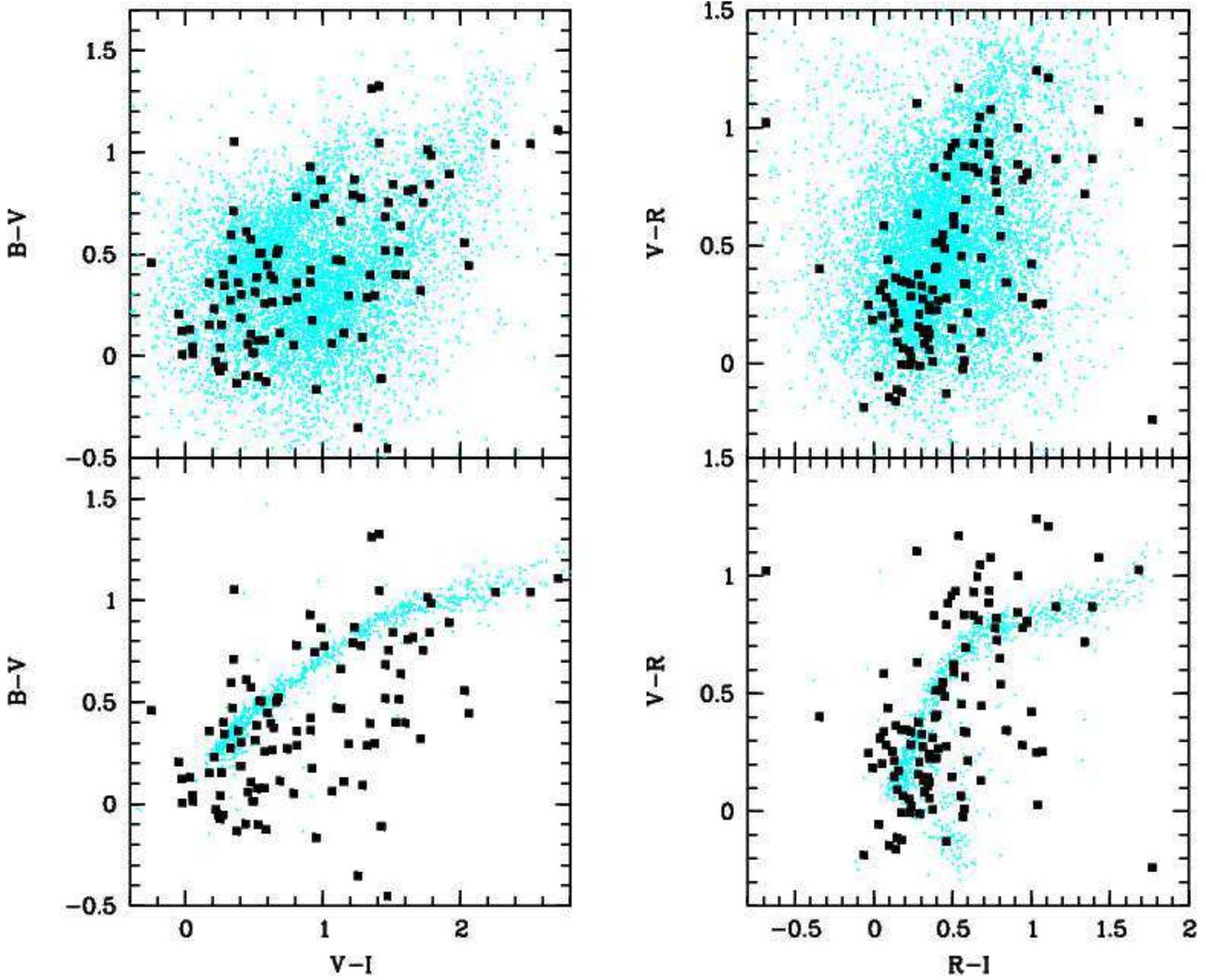}
  \caption{Optical colors for galaxies (top panels) and stars (bottom
    panels) in the field XMM-06. The black squares mark X-ray sources
    in this field with matches to the optical catalogs in all four
    passbands.}
  \label{fig:colors06}
\end{figure*}

In addition, from this preliminary analysis the following conclusions
can be drawn: (1) about 39\% of the X-ray sources have no associated
optical source within 2\arcsec. This optical identification
completeness is comparable with that found by
\citet{2005ApJS..156...35E} in a similar study of X-ray source
samples, but there may also be small contributions from sources with
larger offsets than allowed by the adopted search radius,
contamination by spurious X-ray sources, and random matches with the
optical catalogs; (2) over 50\% of the $m=1$ sources are detected in
all the other bands available, within 1\arcsec; (3) correlations which
occur at search radii greater than 2\arcsec\ are most likely random
correlations with the comparably dense optical catalogs; (4) all
sources which are detected in $RI$ are also detected in $BV$,
indicating that the optical counterparts of the X-ray sources are not
excessively red or, even if they are red, the blue images are
sufficiently deep to detect them; (5) the small number of X-ray
sources matched with objects in the $B$- and $V$-band catalogs without
matches in the $R$-band catalog suggests that we are also not dealing
with excessively blue objects. Figure~\ref{fig:colors06} shows
color-color diagrams for the field XMM-06 for stars and galaxies in
the field and compares it with the optical colors of X-ray sources
matching objects in the optical catalogs. Stars and galaxies were
selected using the SExtractor CLASS\_STAR classifier with the cuts
made at $\mathrm{CLASS\_STAR}<0.1$ and $\mathrm{CLASS\_STAR>0.99}$ for
galaxies and stars, respectively. These diagrams show that, as one
would expect, no specific sub-population of stars or galaxies can be
identified with the X-ray sources.

\section{Summary}
\label{sec:summary}
This paper describes the data products -- reduced and stacked images
as well as science-grade catalogs extracted from the latter --
produced and released for the XMM-Newton follow-up survey performed
with WFI at the ESO/MPG-2.2m telescope as part of the ESO Imaging
Survey project. The survey was carried out as a collaboration between
the EIS, XMM-Newton-SSC and IAEF-Bonn groups. At the time of writing 15 WFI
fields (3.75~square degrees) have been observed for this survey of
which 12 were released in the fall of 2004, with corrections to the
weight maps in July 2005, and are described in this
paper. For the 8 fields at high galactic latitude catalogs are also
presented.

The images were reduced employing the EIS/MVM image processing library
and photometrically calibrated using the EIS data reduction
system. The EIS system was also used to produce more advanced survey
products (stacks and catalogs), to assess their quality, and to make
them publicly available via the web, together with comprehensive
product logs. The quality of the data products reported in the logs is
based on the comparison of different statistical measures such as
galaxy and star number counts and the locus of stars in color-color
diagrams with results obtained in previous works as well as
predictions of theoretical models calibrated by independent
studies. These diagnostics are regularly produced by the system
forming an integral part of it.

In the particular case of this survey, a number of frames have been
reduced by both EIS and the Bonn group, using independent software
thus allowing a direct comparison of the resulting images and catalogs
to be made. From this comparison one finds that the position of the
sources extracted from images produced for the same field/filter
combination by the different pipelines are in excellent agreement with
a mean offset of $\sim$ 20~mas and a standard deviation of $\sim$
50~mas. Comparison of the magnitudes of the extracted sources shows
that in general the mean offset is $\lsim 0.05$~mag, consistent with
the estimated error of the photometric calibration of about
$0.08$~mag. Cases with larger deviations were investigated further and
the problem with the two most extreme cases were found to be unrelated
to the calibration procedure. Instead, it demonstrates the need for
the implementation of additional procedures to cope with the specific
situation encountered and the need for a better calibration plan.
This discussion illustrates a couple of important points. First, that
while an automatic process is prone to errors in dealing with extreme
but rare situations, reductions carried out with human
intervention are prone to random errors which can never be eliminated.
Second, more robust procedures can always be added or existing ones
tuned to deal with exceptions once they are found. However, as always
when dealing with automatic reduction of large volumes of data, the
real issue is to decide on the trade-off between coping with these
rare exceptions and the speed of the process and margin of failures
one is willing to accept.

Finally, a comparison of the PSF distortions suggests that some
improvement could be achieved by requiring the EIS/MVM to impose an
additional constraint on the astrometric solution to improve the
\emph{internal} registration. As mentioned earlier this can be
achieved by imposing the geometrical constraint that the CCDs form a
mosaic.

Preliminary catalogs were also extracted from the available X-ray
images and cross-correlated with the source lists produced from the
$R$-band images. From this analysis one finds that about 61\% of the
X-ray sources have an optical counterpart within 2\arcsec, most of
which are unique. Out of these about 70\% are detected in all the
available passbands. Combined, these results indicate that the adopted
observing strategy successfully yields the expected results of
producing a large population of X-ray sources ($\sim$ 300) with
photometric information in four passbands, therefore enabling a
tentative classification and redshift estimation, sufficiently faint
to require follow-up observations with the VLT.

The present paper is one in the series presenting the
results of a variety of optical/infrared surveys carried out by the
EIS project.

\begin{acknowledgement}
 The results presented in this paper are partly based on observations
  obtained with XMM-Newton, an ESA science mission with instruments
  and contributions directly funded by ESA Member States and
  NASA. This paper was supported in part by the German DLR under
  contract number 50OX0201. JPD was supported by the EIS visitor
  programme, by the German Ministry for Science and Education (BMBF)
  through DESY under the project 05AE2PDA/8, and by the Deutsche
  Forschungsgemeinschaft under the project SCHN 342/3--1. LFO
  acknowledges financial support from the Carlsberg Foundation, the
  Danish Natural Science Research Council and the Poincar\'e
  Fellowship program at Observatoire de la C\^ote d'Azur.
 
\end{acknowledgement}

\bibliographystyle{aa}
\bibliography{3785}

\Online
\appendix

\section{Comments on the individual fields}
\label{sec:field_desc}
Below a broad overview of the 12 fields discussed in the present paper
is given. It includes a description of their nature and the original
target of the X-ray observations. Details about the exposure time per camera
per observation are given on the web-page.

\begin{enumerate}
\item \textbf{XMM-01/RX J0925.7$-$4758} -- The original target was the
  ROSAT-discovered, galactic supersoft X-ray binary
  \object{RX~J0925.7$-$4758}, also known as \object{MR~Vel}
  \citep{1994A&A...284..827M}. The X-ray image is a superposition of
  two medium deep ($\sim60$~ks) observations with EPIC-PN in large
  window, EPIC-MOS1 in full frame, and EPIC-MOS2 in small window mode.
  This results in a reasonably deep ($\sim 115$~ks) exposure in the
  common PN and MOS area (less than half of the field of view). 
  
\item \textbf{XMM-02/RX J0720.4$-$3125} -- Targeting the
  ROSAT-discovered isolated neutron star \object{RX~J0720.4$-$3125}
  \citep{1997A&A...326..662H}, the X-ray image is a superposition of
  three medium deep observations with all cameras in full frame. The
  bright target of the observation causes visible out-of-time events
  (OOT) in the X-ray images leading to bright stripes through the
  target. These stripes are fixed in detector coordinates, hence they
  do not coincide in sky coordinates resulting in three different OOT
  stripes. 

\item \textbf{XMM-03/HE 1104$-$1805} -- The primary target of this XMM-Newton 
  observation was the double-lensed quasar \object{HE~1104$-$1805}
  \citep{1993A&A...278L..15W}. This is an empty field at high galactic
  latitude. The X-ray image is a combined image of the three EPIC
  cameras during the single observation of the target. Bad space
  weather resulted in a loss of $\sim$65\% of the total exposure and a rather
  shallow resulting image ($\sim$10~ks). 
  
\item \textbf{XMM-04/MS1054.4$-$0321} -- The galaxy cluster
  \object{MS~1054.4$-$0321} at redshift of $z=0.83$ is the most
  distant cluster in the Einstein Medium Sensitive Survey and among
  the most massive ones \citep{1990ApJS...72..567G,
    1991ApJS...76..813S, 1994ApJS...94..583G}. The X-ray image is a
  combined image of the three EPIC cameras of the single observation of
  the target, all taken in full frame mode. 
  
\item \textbf{XMM-05/BPM 16274} -- The white dwarf \object{BPM~16274}
  is at high galactic latitude. This field is used for calibrations of
  the optical monitor on-board the XMM-Newton satellite. The X-ray
  image is a superposition of six observations in full frame mode of
  all three cameras. The archive contains already many more data sets with
  other camera settings and the observations in this field are ongoing.
  
\item \textbf{XMM-06/RX J0505.3$-$2849} -- This is a field at high
  galactic latitude. The galaxy cluster \object{RX~J0505.3$-$2849} in
  the center was detected as part of the SHARC survey
  \citep{2003MNRAS.341.1093B} at a redshift of $z=0.509$. The X-ray
  image is the superposition of all three EPIC cameras, which were all
  operated in full frame mode. Only the PN suffered significantly from
  high background, resulting in a medium deep X-ray image
  ($\sim$45~ks). 
  
\item \textbf{XMM-07/LBQS 2212$-$1759} -- In order to avoid potential
  damage from the Leonid meteors when XMM intersects their
  trajectories annually, the spacecraft is oriented in the anti-Leonid
  direction for safety reasons. The field then repeatedly chosen for
  observation was centered on the $z=2.217$ quasar
  \object{LBQS~2212$-$1759} \citep{2001AJ....121.2843B}. The X-ray
  image is a superposition of five observations with all cameras in
  full frame mode. Enhanced background affected less than 15\% of the
  observations resulting in a truly deep field with more than 200~ks
  net exposure in all three cameras. The nominal target of the
  observations was not discovered, the image is thus dominated by the
  serendipitous source content. The most prominent source is a new
  cluster of galaxies close to the center of the field.
  
\item \textbf{XMM-08/NGC 4666} -- The target of this field was the
  almost edge-on spiral galaxy \object{NGC~4666} which dominates the
  center of the field. Although the target was optically and X-ray
  extended, the target was included in the public survey since the
  target blocks only about 5\% of the field of view. The X-ray image
  is a combined image of the three EPIC cameras of the single
  observation of the target, all taken in full frame mode. 
  
\item \textbf{XMM-09/QSO B1246$-$057} -- The target was the broad
  absorption line quasar \object{B1246$-$057}. The XMM-Newton
  observation of the target was published by
  \citet{2003AJ....126.1159G}. The X-ray image is a combined image of
  the three EPIC cameras of the single observation of the target, all
  taken in full frame mode. The field is unusual, about one quarter 
  of the field in the NE corner is almost devoid of X-ray sources. 
  The dominant point-like object is the X-ray
  counterpart of the Algol binary \object{HD~111487}
  (\object{RBS~1165}).
  
\item \textbf{XMM-10/PB 5062} -- \object{PB~5062} (also known as
  \object{QSO~B2202$-$0209}) is a quasar at redshift $z=1.77$
  \citep{1998A&A...330..108Y}. The X-ray image is a combined image of
  the three EPIC cameras of the single observation of the target, all
  taken in full frame mode. 
  
\item \textbf{XMM-11/Sgr A} -- The target of this field is the center
  of our Milky Way. The X-ray image is a superposition of five
  observations with all cameras in full frame mode. The observations
  were severely affected by background flares resulting in a loss of
  more than 50\% of the observation time. Nevertheless, repeated
  observations of the field resulted in a medium deep exposure with
  more than 40~ks exposure in all three cameras. The image is
  dominated by diffuse emission from the galactic center region. In
  addition there are two classes of point-sources, heavily absorbed,
  i.e. likely background sources with hard X-ray spectra and
  foreground sources with soft X-ray spectra.
  
\item \textbf{XMM-12/\object{WR 46}} -- This object also known as
  \object{HD~104994} is a Wolf-Rayet star. The X-ray image is a combined
  image of the three EPIC cameras of the single observation of the
  target, all taken in full frame mode. Data loss due to enhanced
  background is insignificant resulting in a medium deep field
  with more than 70~ks exposure in each camera. 
\end{enumerate}

\section{Photometry}
\label{sec:photometry}
The photometric calibration is done in a fully automated way by the
EIS Data Reduction System (da Costa et al., in preparation)
calibrating all data to the Vega magnitude system. The photometric
pipeline extracts catalogs from the standard star fields, and measures
fluxes at different apertures to allow for a growth curve inspection.
The positions and flux measurements for each object are
cross-correlated with those corresponding to known standard stars as
stored in a database. The matched standard stars are then used to
determine the photometric zeropoint, extinction and color term from a
linear fit. Finally, the zeropoint for each science frame is derived
from the linear fit derived for the night of observation.

In the case of optical observations, as in this paper, the
calibrations have been based on observations of
\citet{1992AJ....104..340L} standard star fields. The magnitudes were
measured in 6\arcsec apertures, which from monitoring the growth curve
of all measured stars proved to be adequate. The linear fits had from
one to three free parameters depending on the available airmass and
color coverage provided by the calibration plan. In cases where the
airmass and/or color coverage is insufficient, pre-specified values
for the extinction and color term, either determined from other
solutions or from theoretical models, are used.

The photometric pipeline computes photometric parameters for all
possible types of fits (one to three free parameters) and chooses the
solution with the smallest rms to be the \emph{best solution} for the
night. A night is considered photometric if this scatter is less than
a pre-defined value, which at the present time is taken to be 0.1~mag.
If none of the solutions satisfies this criterion and/or the solution
found yields unrealistic results (e.g. negative extinction) then the
night is considered non-photometric and a default value for the
zeropoint is adopted and its error set to $-1$. For nights without
observation of standard star fields, a default zeropoint and an error
of $-2$ are assigned to the image. Finally, during quality assessment
of the data, calibrated images with zeropoints that deviate
significantly from a reference value have the zeropoint in the header
changed to a default value and its error set to $-3$. When a
default value is assigned all images will have the same zeropoint in
the header regardless of the airmass at which they were observed. For
homogeneity, the default value normally adopted is the median of the
zeropoints reported in the trend analysis kept by either the telescope
team (depending on the instrument) or the internal EIS database. In
the case of WFI, only one solution is currently reported on the WFI
WWW pages.\footnote
{\url{http://www.ls.eso.org/lasilla/sciops/2p2/E2p2M/WFI/zeropoints/}}

The derived photometric solutions were used to calibrate the reduced
images. The zeropoint in the header of a reduced image is given by
\begin{equation}
  \label{eq:1}
  ZP = ZP' - kX
\end{equation}
where $ZP'$ is the zeropoint at zero airmass (determined from the
linear fit as described above), $k$ is the extinction coefficient and $X$
is the airmass at which the frame was observed. Using this definition
the Vega magnitude can computed by 
\begin{equation}
  \label{eq:2}
  m_\mathrm{Vega} = -2.5\log(f) + ZP
\end{equation}
where $f$ is the flux in number of counts directly measurable on
the image (note that the reduced images are normalized to 1~s).

\section{Image stacks}
\label{sec:image-stacks}
Most fields are covered by more than one reduced image. These are
co-added to create the final image product. The following steps are
involved in the creation of a final stacked image:

\begin{itemize}
\item Grouping: reduced images are grouped into \emph {stacking
    blocks} (SBs) according to position (with a minimum distance
  between centers of 0.25 times the field-of-view), and filter.
  
\item Validation: as the co-addition is carried out in pixel space it
  is required that all images have been warped to the same reference
  grid. Therefore, images in a SB must share the same reference grid
  (projection, reference position, pixel scale and orientation). In
  addition, the images contributing to an SB, or their original RBs,
  are checked to ensure that they appear only once in the SB. Images
  in the SB are also checked to ensure that their flux scale has been
  properly normalized to 1~s, and if not they are re-normalized
  accordingly.
  
\item Constraints: images in an SB are checked to ensure that they
  meet certain constraints, for instance on the value of seeing, the
  rms of the PSF distortion and grade. In addition, the contributing
  images to an SB, or their original RBs, are checked to prevent the
  repetition of raw exposures. Images not satisfying the constraints
  are discarded from the SB.

  \item PSF homogenization: all images in the SB are convolved by a
  Gaussian with a FWHM corresponding to the the largest value computed
  for the contributing images.

  \item Flux scale determination: before the images can be co-added,
  which is done using a weighted mean procedure, it is necessary to
  have all images at the same flux level. This is done by scaling
  photometric frames to zero airmass using the extinction coefficients
  from the photometric solutions. The non-photometric frames are
  scaled to the zero airmass level of the photometric frames (see
  below)

  \item Co-addition: images in a validated SB are co-added using a
  weighted mean. The weight images reflect the exposure times of their
  associated science image. Therefore, the weight of each image is a
  combination of exposure time and noise. However, the weight images
  also contain information about the location of bad pixels and masks
  which are set to zero weight and thus do not contribute to the final
  image. In the co-addition process, performed by the routine
  add-mosaic of the EIS/MVM library, a sigma-clipping procedure is
  employed to remove cosmic ray hits.
\end{itemize}

The image co-addition uses a weighted mean combination. The weighting
is done with the weight images produced by the image reduction.
Additionally, a thresholding procedure is employed to remove cosmic
rays. In general, the use of weight maps and cosmic ray hits removal
produces very clean final images.

As mentioned above, all images in an SB have to be scaled to a common
flux level before the co-addition to assure the photometric
calibration of the final, stacked image. For this rescaling, one has
to consider two cases:

\begin{enumerate}
\item \textbf{SB with at least one photometric frame} -- In this case
  the reference flux level is obtained from the combination of all
  contributing photometric frames. First, all photometrically
  calibrated frames are scaled to the flux level that would have been
  obtained at zero airmass. This is done using the extinction
  coefficients of their respective photometric solutions. Second, by
  computing a weighted average of all scaled images a reference image
  is created. For the final co-addition all input images (photometric
  and non-photometric) are scaled to the flux level of the reference
  image. The scaling factors are determined by comparing object
  magnitudes. Naturally, in cases where only one photometric frame is
  available, this one is used as the reference.
  
\item \textbf{SB with no photometric frames} -- Occasionally, none of
  the reduced images in an SB are photometrically calibrated (e.g.  no
  standards were observed in the night, observations in
  non-photometric nights). In such cases, all images are scaled to the
  flux level of an arbitrarily chosen reference image from the SB. The
  zeropoint of the final stacked image is then the zeropoint of the
  adopted reference image. The zeropoint of the output image is that
  of the arbitrarily chosen reference. These frames can be identified
  by the negative zeropoint error assigned to these cases, as
  described above.

\end{enumerate}

\section{Catalog production}
\label{sec:catalogs}
The final processing step in the EIS Data Reduction system is the
creation of source catalogs by standardized procedures, resulting in
catalogs containing enough information to be directly usable for
scientific applications. Here the production of catalogs for deep,
sparse fields is discussed. The production of catalogs for crowded
fields, which are not included in this release, will be described in
Zaggia et al. (2005, in preparation).

The EIS catalog production is based on
SExtractor \citep{1996A&AS..117..393B} and a common configuration file
for all catalogs with a minimum number of adjustments to be made for
individual images. For each image the appropriate values for the
seeing and magnitude zeropoints as well as the weight-map associated
to each image are used. Other parameters are the same for all
catalogs.

The catalog production starts with a very low $S/N$
catalog, which contains a large number of spurious objects. To produce
a science-grade catalog a number of steps are taken. First, the
catalog is pruned for objects with a $S/N$ (determined from the
MAGERR\_AUTO) below a user-defined level, which was set to 5
for this release. The object magnitudes are converted to the AB system
according to the response function of the optical system and corrected
for galactic extinction. At present, this correction is
applied to the magnitudes of all objects, including stars. To
facilitate the use of the catalogs 14 flags are added for each object
as described below.

The saturation level of the final image stack is difficult to determine
from the input images due to variations in integration time and possibly
seeing. Therefore, the saturation level is determined from the extracted
catalog. The method is based on the FWHM and peak flux of
bright objects. The distribution of the FWHM is determined and
sigma-clipped to exclude bright galaxies from the sample. Among the
remaining objects those with FWHM deviating more than $3\sigma$ are
taken to be the saturated objects. The saturation level is set to the
minimum peak value among these objects. This value is used to set the
saturation flag. 

To be able to remove objects close to bright stars masks can be
created in two ways: by an automatic routine or by hand using a Skycat
plug-in. The automatic masking adds masks around saturated objects as
well as around objects brighter than a user-specified magnitude. The
size of the mask scales with the major-axis of the object, as computed
by SExtractor. The scaling factor is specified by the user. The
adopted shape of the masks is a square with one of the diagonals
oriented North-South in an attempt to mask the diffraction spikes. The
parameters used in the mask definition are reported in the product
log. The positions of all masks are reported in the \texttt{MASKS}
table in the catalog.

Note that except for objects with $S/N$ less than that required, no
object is removed from the catalog. If necessary, more objects can be
pruned by the user according to the flags described below. In
addition, the default magnitude system adopted for the objects can be
changed using the information available in the \texttt{FIELDS} table.

The catalogs produced by the EIS Data Reduction system are in
FITS format, based on the ``Leiden Data Center'' (LDAC) convention
originally adopted by the DENIS project and later expanded in the
course of the EIS project. It currently consists of a FITS header and
the following tables: \texttt{FIELDS}, \texttt{OBJECTS},
\texttt{MASK}, and \texttt{FILTER}.

The \texttt{FIELDS} table contains general information for all objects
in the catalogs. It consists of 109 columns including: (1) basic
information set by the LDAC library; (2) keywords taken from the FITS
header of the image from which the catalog was extracted; (3) the main
SExtractor configuration parameters used; and (4) information computed
by the EIS Data Reduction system. The latter includes, for instance:
\begin{itemize}
\item the diameter of the 10 apertures used for aperture magnitudes,
  ranging from 1\arcsec\ to 5\arcsec\ in steps of $0\farcs5$ and a
  large aperture of 10\arcsec;
  
\item the WCS coordinates of the corners of the original image and of
  the trimmed area;
  
\item the extinction correction. This is computed as the average of
  the value of the extinction in cells of $3\arcmin\times3\arcmin$,
  distributed over the trimmed image;
  
\item the value added to the original magnitude of the extracted
  objects in the Vega system to produce the reported magnitudes in the
  catalog in the AB system;
  
\item an estimate of the fudge factor used to multiply the errors
  reported by SExtractor to correct for the correlated noise
  introduced by the re-sampling kernel;

\item total and trimmed areas.

\end{itemize}
Some of the information contained in the \texttt{FIELDS} table is also
available in the \emph{Product Logs} which are available on the EIS
XMM-Newton follow-up survey release pages.

The \texttt{OBJECTS} table reports the parameters characterizing the
individual extracted objects as computed by SExtractor. It has 69
columns, some being vectors (e.g. aperture magnitudes), describing the
main geometric and photometric properties of the objects. The
parameters were chosen as a compromise between the total number of
parameters and the most frequently requested parameters from survey
product users. The choice of apertures and the flags defined are the
result of suggestions made by users of EIS data products. In addition
to the SExtractor flag, which are described in the SExtractor manual,
14 other flags have been defined to facilitate the filtering of the
catalogs. These are
\begin{itemize}
\item \texttt{FLAG\_SEX1}--\texttt{FLAG\_SEX128} -- 8 flags
  individually representing the various SExtractor flag components;

\item \texttt{FLAG\_SAT} -- set to 1 if the object is saturated;

\item \texttt{FLAG\_TRIM} -- set to 1 if the object is inside a trimmed area;

\item \texttt{FLAG\_MASK} -- set to 1 if the object is inside a masked area;

\item \texttt{EISFLAG} -- sum of \texttt{FLAG\_TRIM} and \texttt{FLAG\_MASK}

\item \texttt{FLAG\_STATE} -- 1 if any of the above flags are set

\item \texttt{FLAG\_STAR} -- 1 if star, 0 if galaxy, based on
  SExtractor's CLASS\_STAR parameter. The value used for separation
  and the magnitude down to which a classification was attempted are
  reported in the product log.
\end{itemize}

The \texttt{MASK} table gives the number and coordinates of the vertices
of both automatically created masks as well as those drawn by hand
using a Skycat plug-in.

The \texttt{FILTER} table gives the filter transmission curve and their
convolution with the optical system response function.

\end{document}